\documentclass[
amsmath, %
floatfix, %
twocolumn, %
reprint, %
superscriptaddress, %
prb, %
aps, %
citeautoscript, %
final, %
]{revtex4-2} 
\renewcommand{\thispagestyle}[1]{}
\usepackage[]{newtxtext} 
\usepackage[subscriptcorrection,nosymbolsc,smallerops,bigdelims]{newtxmath} 
\DeclareMathAlphabet{\mathcal}{OMS}{cmsy}{m}{n} 
\DeclareMathAlphabet{\mathbcal}{OMS}{cmsy}{b}{n} 
\usepackage{bm}
\newcommand*{\oncite}[1]{Ref.~[\onlinecite{#1}]} 

\usepackage[utf8]{inputenc}
\usepackage[T1]{fontenc}

\usepackage[]{graphicx}
\usepackage{latexsym}

\usepackage{wasysym}

\usepackage{color}
\usepackage{mathtools}
\usepackage[section]{placeins}
\usepackage[]{xcolor}
\usepackage{bm}
\usepackage{siunitx}
\usepackage{hyperref}
\hypersetup{
	colorlinks,
	linkcolor={blue!90!black},
	citecolor={blue!90!black},
	urlcolor=	{blue!90!black}
}
\let\oldeqref\eqref
\renewcommand*{\eqref}[1]{%
	Eq.~\hyperref[eq:#1]{\oldeqref{eq:#1}}%
}

\usepackage{floatrow}
\newcommand{\mr}[1]{\mathrm{#1}}
\newcommand{\ee}{\mr{e}}

\newcommand{\rr}{\bm{r}}

\mathchardef\mhyphen="2D

\DeclarePairedDelimiter\lr{\lparen}{\rparen}
\DeclarePairedDelimiter\Lr{\lbrack}{\rbrack}
\DeclarePairedDelimiter\LR{\lbrace}{\rbrace}
\DeclarePairedDelimiter\abs{\lvert}{\rvert}

\DeclarePairedDelimiterX{\comm}[2]{\lbrack}{\rbrack}{#1, #2}
\DeclarePairedDelimiter\ket{\lvert}{\rangle}

\DeclarePairedDelimiterX{\braket}[2]{\langle}{\rangle}{#1\delimsize\vert #2}
\DeclarePairedDelimiterX{\ketbra}[2]{\rvert}{\lvert}{#1 \delimsize\rangle\!\delimsize\langle #2}
\DeclarePairedDelimiterX{\matrixel}[3]{\langle}{\rangle}{#1 \delimsize\vert #2 \delimsize\vert #3}

\newcommand{\figref}[1]{Fig.~\ref{fig:#1}}
\newcommand{\subfigref}[2]{Fig.~\hyperref[fig:#1]{\ref*{fig:#1}(#2)}}
\newcommand{\subfigsref}[3]{Figs.~\hyperref[fig:#1]{\ref*{fig:#1}(#2)}-\hyperref[fig:#1]{\ref*{fig:#1}(#3)}}

\newcommand{\newtext}[1]{#1}

\usepackage[low-sup]{subdepth}
\usepackage{ragged2e}

\definecolor{cbred}{HTML}{e31a1c}
\definecolor{cbgreen}{HTML}{33a02c}
\definecolor{cbblue}{HTML}{176aa7}
\definecolor{cborange}{HTML}{ff7f00}
\definecolor{cbviolet}{HTML}{6a3d9a}
\definecolor{cbbrown}{HTML}{b15928}

\definecolor{cblred}{HTML}{fb9a99}
\definecolor{cblgreen}{HTML}{b2df8a}
\definecolor{cblblue}{HTML}{a6cee3}
\definecolor{cblorange}{HTML}{fdbf6f}
\definecolor{cblviolet}{HTML}{cab2d6}
\definecolor{cblbrown}{HTML}{ffff99}

\usepackage[activate={true,nocompatibility},final,tracking=alltext,kerning=true,spacing=true,protrusion=true,factor=1100,stretch=4,shrink=6,selected=true ,letterspace=-0]{microtype}

\begin{document}

\title{Bardeen's tunneling theory applied to \newtext{intraorbital and interorbital} hopping integrals between dopants in silicon}

\author{Micha\l\ Gawe{\l}czyk}
\email{michal.gawelczyk@pwr.edu.pl}
\affiliation{Institute of Physics, Faculty of Physics, Astronomy and Informatics, Nicolaus Copernicus University in Toru\'n, Grudzi\k{a}dzka 5, 87-100 Toru\'n, Poland}
\affiliation{Department of Theoretical Physics, Faculty of Fundamental Problems of Technology, Wroc\l{}aw University of Science and Technology, 50-370 Wroc\l{}aw, Poland}

\author{Micha{\l} Zieli\'nski}
\affiliation{Institute of Physics, Faculty of Physics, Astronomy and Informatics, Nicolaus Copernicus University in Toru\'n, Grudzi\k{a}dzka 5, 87-100 Toru\'n, Poland}

\begin{abstract}
	We utilize Bardeen's tunneling theory to calculate intra- and interorbital hopping integrals between phosphorus donors in silicon using known orbital wave functions.
	\newtext{While the two-donor problem can be solved directly, the knowledge of hoppings for various pairs of orbitals is essential for constructing multi-orbital Hubbard models for chains and arrays of donors.
	To assure applicability to long-range potentials, we rederive Bardeen's formula for the matrix element without assuming non-overlapping potentials.
	Moreover, we find a correction to the original expression allowing us to use it at short distances.}
	We also show that accurate calculation of the lowest donor-pair eigenstates is possible based on these tunnel couplings, and we characterize the obtained states.
	The results are in satisfactory quantitative agreement with those obtained with the standard H\"uckel tight-binding method.
	The calculation relies solely on the wave functions in the barrier region and does not explicitly involve donor or lattice potentials,
	\newtext{which has practical advantages.
	We find that neglecting the central correction potential in the standard method may lead to qualitatively incorrect results, while its explicit inclusion raises severe numerical problems, as it is contained in a tiny volume.
	In contrast, using wave functions obtained with this correction in the proposed method does not raise such issues.}
	Nominally, the computational cost of the method is to calculate a double integral along the plane that separates donors. 
	For donor separation in directions where valley interference leads to oscillatory behavior, additional averaging over the position of the integration plane is needed.
	Despite this, the presented approach offers a competitive computational cost as compared to the standard one.
	This work may be regarded as a benchmark of a promising method for calculating hopping integrals in lattice models with known or postulated orbital wave functions.
	
\end{abstract}
\maketitle
	
\section{Introduction}
	After initial proposals of silicon-based quantum computer architectures \cite{KaneN1998,VrijenPRA2000,HollenbergPRB2004}, and subsequent efforts \cite{OBrienPRB2001,JarrydN2013,DehollainPRL2014,GonzalezNL2014}, atomically precise phosphorus donor placement in silicon became achievable \cite{FuechsleNN2012,BuchNC2013,WyrickAFM2019,WangCP2020,ZwanenburgRoMP2013,AlipourJVS2022}.
	This development finally made it possible to demonstrate a quantum simulator of the extended Hubbard model based on a 2-D donor array \cite{Wang2021}.
	In parallel to these technological and experimental advances, the development of a theoretical description of donor levels \cite{KohnPR1955} continues, with the effective mass theory (EMT) \cite{KlymenkoJPCM2014,SaraivaJPCM2015,GamblePRB2015} and atomistic tight-binding \cite{MencheroPRB1999,MartinsPRB2005,KlimeckCMiES2002,RahmanNano2011,TankasalaPRB2022} approaches giving results well conforming to the experimental findings.
	Recently, EMT allowed for a detailed calculation of excited two-electron states of a donor pair and their optical spectra \cite{WuPRB2021}.
	Apart from the correct prediction of donor orbital energies \cite{JagannathPRB1981,MayurPRB1993}, detailed modeling of wave functions has been achieved \cite{SalfiNC2014,GamblePRB2015}, which made possible the calculation of parameters \cite{KoillerPRL2001,XuPRB2005,WellardPRB2005,QiuziPRB2010,LePRB2017a,DuskoNPJQI2018} such as hopping and exchange integrals, on-site energy, and Coulomb repulsion, i.e., matrix elements of one- and two-body operators.
	These together allow for a two-way correspondence between donor chains or arrays and lattice models of the Fermi-Hubbard type \cite{DuskoNPJQI2018}.
	With this, an interesting path of observing and describing the onset of collective effects in small many-body systems opens \cite{TownsendPRB2021}.

	The most basic parameter of such models is the hopping integral (tunnel coupling) $t$.
	It is also the one that is not easily accessible experimentally, which makes its modeling of great importance.
	The tunnel coupling can be found as half of the splitting between the bonding and antibonding states of a donor pair.
	The most accurate but also computationally challenging way of evaluating it would be to directly calculate the two-donor states in a multi-orbital model, separately for every donor displacement.
	For a more feasible calculation, the H\"uckel tight-binding theory \cite{AshcroftBook1976} can be successfully utilized to determine $t$ based on the known single donor ground state and $1/r$ donor Coulomb potential \cite{LePRB2017a}.
	
	Here, we propose another method and evaluate the hopping integrals utilizing Bardeen's transfer Hamiltonian theory, originally derived for the problem of electron tunneling between many-body eigenstates of two regions separated by a potential barrier \cite{BardeenPRL1961}.
	It provides an expression for the matrix element involving only the initial and final eigenstates from the two regions.
	Most importantly, no specific knowledge about the potential barrier is needed except that it is wide enough to fulfill the assumptions of the theory.
	This formulation made it particularly suitable for calculating tunneling currents in various types of junctions.
	After adaptation \cite{TersoffPRL1983,TersoffPRB1985} and further development \cite{ChenPRB1990}, it became a standard calculation method in scanning tunneling microscopy (STM), both for theoretical simulation \cite{DrakovaRPP2001} and for the translation of the measured current into atomistically resolved scans \cite{TsukadaASS1994}, i.e., for the interpretation of STM data.
	
	Using this method to calculate hopping between donors, we take advantage of the fact that it only relies on wave functions.
	Thanks to this, it may be applied, based on known or postulated functions, \newtext{to various systems, including, e.g., arrays of quantum dots, other defect systems, or systems of nanoscale thickness with dielectric mismatch \cite{RyuNanoscale2013}, based on a single solution of the one-site problem followed by a computationally cheap application of the method to pairs or other arrangements of sites.
	The method can be thus used} without explicitly knowing or expressing the potential of the donor (site) and the surrounding lattice.
	\newtext{In the case of donors, where the ionic $1/r$ potential has to be augmented with a central cell correction, a purely wave-function-based method turns out to have practical advantages.
	We show that for the standard method applied to the given system, there are cases in which pure $1/r$ potential may give falsely vanishing results, while the explicit inclusion of the correction potential in the integration involved turns out to be numerically troublesome due to its small spatial extension.
	In contrast, the integration of wave functions that intrinsically contain central cell correction effects in the proposed method is not so problematic.}
	
	\newtext{For donor arrays with lattice spacings below $\sim6$~nm, the need to consider at least a second orbital in Hubbard-like models arises due to orbital mixing \cite{SaraivaJPCM2015,GamblePRB2015,TankasalaPRB2022}.
	This may allow studying phenomena like the orbital-selective Mott transition \cite{LiebschPRL2005}.
	While full solutions of the two-donor problem \cite{GamblePRB2015,TankasalaPRB2022} are available, in this range of distances, the splitting of the two lowest eigenstates from such a solution loses the meaning of hopping.
	Only the lowest-orbital "constituent" hoppings can be found in the literature \cite{LePRB2017a}, and those among different orbitals cannot be extracted from a direct solution of the problem.}

	Thus, we calculate here all nonvanishing hopping integrals between the six ground-state orbitals at each of the donors, i.e., both intra- and interorbital tunnel couplings.
	In the case of vanishing integrals, we give a symmetry-based justification.
	For comparison, we also use the standard method.
	In most cases, the results are in good agreement, which we consider a mutual confirmation since both methods are intrinsically approximate.
	Next, we use all these integrals to form a two-donor Hamiltonian and find its eigenstates.
	These are in good agreement with a full two-donor EMT calculation \cite{GamblePRB2015}.
	We characterize the eigenstates showing their composition in terms of valleys and single-donor orbitals. 
	
	\newtext{As the original derivation of Bardeen's formula for the matrix element assumes non-overlapping potentials, which is not fulfilled here, we present a derivation that does not rely on this assumption.
	Moreover, while rederiving the expression, we find a correction that extends the range of its applicability at short distances and hence for significant wave-function overlaps.}
	
	Finally, we evaluate the computational cost of both methods, which turns out to be similar.
	Based on this and on the conceptual and practical advantages of the proposed method, we find it to be prospective for the calculation of hopping integrals based on known or postulated wave functions in various physical systems.
	
	In the following, we first introduce the system and its theoretical model in Sec.~\ref{sec:system}, then present and discuss the general results in Sec.~\ref{sec:results}. \newtext{Next, in Sec.~\ref{sec:corr}, we derive a correction to Bardeen's formula and show its results.} Finally, we conclude the study in Sec.~\ref{sec:conclusions}. \newtext{In Appendix~\ref{sec:huckel}, we describe the standard method used by us for comparison.} More technical calculation details are given in Appendix~\ref{sec:details}, while in Appendix~\ref{sec:calctime}, we estimate the computational cost of the calculation.
	
\section{System and theoretical model}\label{sec:system}
	\begin{figure}[tb]
		\includegraphics[width=0.9\linewidth]{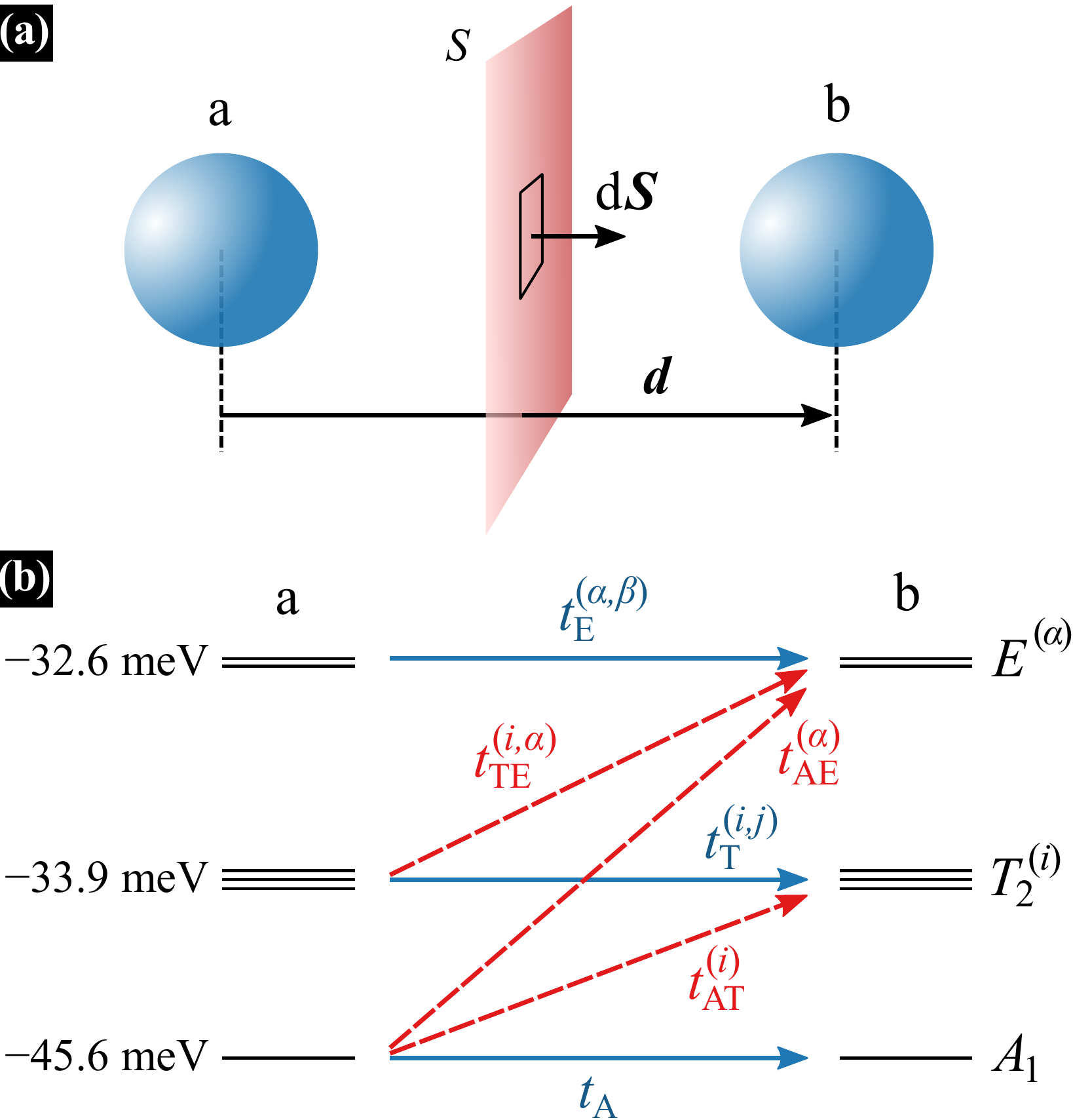} %
		\caption{\label{fig:scheme}(Color online) Schemes. (a) A schematic view of the system with the integration plane drawn. (b) Schematic view of donor levels with relevant intra- (solid blue) and interorbital (dashed red) tunnel couplings marked with arrows. Note that energy splittings in (b) are not to scale, i.e., vertical distances do not reflect real splittings; in particular, artificial spacing in degenerate $T_2$ and $E$ multiplets is introduced to show the number of levels.} %
	\end{figure}
	
	We deal with a system composed of two substitutional phosphorus donors in bulk silicon separated by displacement $\bm{d}$, so located at a distance $d=\abs{\bm{d}}$ apart.
	In \subfigref{scheme}{a}, we schematically present the system.
	Additionally, we show the integration plane and its exemplary element, which will be helpful later. 
	Our aim is to calculate tunnel couplings between various pairs of orbitals localized at two different donors.
	The ground-state manifold contains six states named after their symmetry: $A_1$, three degenerate $\cramped{T_2^{(x)}}$, $T_2^{(y)}$, $T_2^{(z)}$, and two degenerate $E^{(xy)}$, $E^{(z)}$.
	Note that the labels for $E$ orbitals are shorthand, as they transform as $x^2-y^2$, $2z^2-x^2-y^2$, respectively.
	We denote the hopping integrals among pairs of orbitals by $t$ with various indices.
	In the subscript, a single label means a same-type orbital hopping, while two labels are given for a matrix element between different orbital types. Orbital indices for degenerate multiplets are given in the superscript, where $i,j\in\LR{x,y,z}$ and $\alpha,\beta\in\LR{xy,z}$ relate to the $T_2$ and $E$ orbitals, respectively.
	Thus, for instance, $t_{AT}^{(x)} = \matrixel{A_1}{H_{\mathrm{t}}}{\,T_2^{(x)}}$, and $t_{TE}^{(x,z)} = \matrixel{T_2^{(x)}}{H_{\mathrm{t}}}{\,E^{(z)}}$, where $H_{\mathrm{t}}$ is the tunnel coupling Hamiltonian.
	The ladder of orbital levels is shown in \subfigref{scheme}{b} with all unique variants of calculated couplings schematically marked with arrows and labels, which can make the naming scheme easier to follow.
	
	The system is described by the Hamiltonian
	\begin{equation}\label{eq:hamiltonian}
		H = H_{\mathrm{a}} + H_{\mathrm{b}} + H_{\mathrm{t}},
	\end{equation}
	where $H_{\mathrm{a}}$ and $H_{\mathrm{b}}$ are identical except for different positions and are the energies of isolated donors, while the already announced tunnel (or transfer) Hamiltonian $H_{\mathrm{t}}$ is the part responsible for coupling between donors.
	According to Bardeen's theory, matrix elements of $H_{\mathrm{t}}$ can be calculated by integrating the transition probability current operator
	\begin{equation}\label{eq:trans-prob-curr}
		J_{ij}\mkern-1mu\lr{\rr} = \psi_i^{*}\lr*{\bm{r}}\frac{\hbar^2\nabla}{2m}\psi_j\lr*{\bm{r}} - \psi_j\lr*{\bm{r}}\frac{\hbar\nabla}{2m}\psi_i^{*}\lr*{\bm{r}}
	\end{equation}
	over an arbitrary surface $S$ separating the sites given the surface lays in the potential barrier region [see \subfigref{scheme}{a}],
	\begin{align}\label{eq:bardeen}
		\matrixel{i}{H_{\mathrm{t}}}{j} ={}& \frac{\hbar^2}{2} \int_{S} \mathrm{d}\bm{S} \, \Lr*{ \psi_i^{*}\lr*{\bm{r}}\frac{\nabla}{m}\psi_j\lr*{\bm{r}} - \psi_j\lr*{\bm{r}}\frac{\nabla}{m}\psi_i^{*}\lr*{\bm{r}} } \nonumber\\
		&- \lr*{E_i-E_j} \, \int_{V_+} \mathrm{d}^3\bm{r} \, \psi_i^{*}\lr*{\bm{r}}\psi_j\lr*{\bm{r}},
	\end{align} 
	where $\mathrm{d}\bm{S}$ is the surface element, $\psi_i\lr*{\bm{r}}=\braket{\bm{r}}{i}$ is the wave function of the $i$-th donor level, and $E_i$ is its energy. The second term on the right-hand side is a correction for the case of nondegenerate levels \cite{ReittuAJoP1995}, in which $V_+$ means the volume on one side of $S$.
	In \eqref{bardeen}, we deliberately keep the mass $m$ inside the expression, as in the silicon matrix, we need to take into account the anisotropy and valley dependence of the effective mass operator.
	At this point, we may notice that by choosing $S$ to be a plane perpendicular to the donor displacement, $S\perp\bm{d}$, we may simplify the calculation of the first term
	\begin{multline}\label{eq:gradtodx}
		\int_{S} \mathrm{d}\bm{S} \, \Lr*{ \psi_i^{*}\lr*{\bm{r}}\frac{\nabla}{m}\psi_j\lr*{\bm{r}} - \psi_j\lr*{\bm{r}}\frac{\nabla}{m}\psi_i^{*}\lr*{\bm{r}} } \\
		\stackrel{ S\perp\bm{d}}{=}
		\int_S \mathrm{d}s \, \Lr*{ \psi_i^{*}\lr*{\bm{r}}\frac{1}{m} \frac{\partial\psi_j\lr*{\bm{r}}}{\partial x} - \psi_j\lr*{\bm{r}}\frac{1}{m} \frac{\partial\psi_i^{*}\lr*{\bm{r}} }{\partial x} },
	\end{multline}
	where coordinate axes are chosen such that $x$ is along the displacement, $\widehat{\bm{x}}\parallel\bm{d}$, and $\mathrm{d}s = \mathrm{d}y\mathrm{d}z$.
	
	\newtext{The general results in the next section are obtained with \eqref{bardeen} and are in satisfactory agreement with the standard method.
	However, in Sec.~\ref{sec:corr}, we show that the formula can be further corrected to be more accurate at short distances (for high wave function overlaps).}
	
	To evaluate the above integrals, we use the orbital wave functions provided in \oncite{GamblePRB2015}.
	They are calculated within the multivalley effective mass theory \cite{KohnPR1955,LuttingerPR1995,ShindoJPSJ1976} with a symmetry-adapted central cell correction \cite{CastnerPRB2009,GreenmanPRB2013} and were shown to reproduce orbital energies from the experiment \cite{JagannathPRB1981,MayurPRB1993} and lead to position-dependent results quantitatively consistent with those of the atomistic tight-binding method \cite{MencheroPRB1999,MartinsPRB2005,KlimeckCMiES2002}.
	For the latter, in turn, the agreement with measured wave function was demonstrated \cite{SalfiNC2014}.
	Thus, the used wave functions provide a good basis for reliable calculations.
	
	The wave function $\psi\lr{\rr}$ is expressed as
	\begin{equation}
		\psi\lr{\rr} = \sum_{\mu} F_{\mu}\lr{\rr} \, \phi_{\mu}\lr{\rr},
	\end{equation}
	where $\mu\in\LR{-x,+x,-y,+y,-z,+z}$ runs over the six $\bm{k}$-space valleys at $\bm{k}_{\mu} = 0.84\times(2 \pi /a) \,\widehat{\mu}$.
	Here, $\widehat{\mu}$ are the corresponding Cartesian unit vectors, $\widehat{\mu}\in\LR{[100],~[010],~[001]}$, $a=0.54307$~nm is the Si lattice constant, and
	\begin{equation}
		\phi_{\mu}\lr{\rr} = u_{\bm{k}_{\mu}}\lr{\rr} \, \ee^{i\bm{k}_\mu\!\cdot\rr}
		= e^{i\bm{k}_\mu\!\cdot\rr} \sum_{\bm{G}} A^{(\mu)}_{\bm{G}} \ee^{i\bm{G}\cdot\rr}
	\end{equation}
	are the Bloch functions for the respective valley minima with the periodic part $u_{\bm{k}_{\mu}}\lr{\rr}$ expanded in plane waves with $A^{(\mu)}_{\bm{G}}$ being the coefficients, and $\bm{G}$ the reciprocal lattice vectors.
	Bloch functions are weighted by slowly varying envelopes $F_{\mu}\lr{\rr}$,
	\begin{equation}
		F_{\mu}\lr{\rr} = \sum_i B_{\mu,i} F_{\mu,i}\lr{\rr}
	\end{equation}
	that are in turn expanded in a basis of Gaussian envelopes (identical for all valleys upon coordinate permutation) with coefficients $B_{\mu,i}$.
	Details on calculations of wave functions and data allowing for their reproduction can be found in \oncite{GamblePRB2015}.

	Let us now focus on the derivative of $\psi\lr{\rr}$,
	\begin{multline}
		\!\!\!\!\!\lr*{ \frac{1}{m}\frac{\partial}{\partial x} } \,\psi\lr*{\rr}\\
		~~= \sum_{\mu} \Lr*{ \phi_{\mu}\lr{\rr}\, \lr*{ \frac{1}{m}\frac{\partial}{\partial x} }\, F_{\mu}\lr{\rr} + F_{\mu}\lr{\rr} \, \lr*{ \frac{1}{m}\frac{\partial}{\partial x} }\, \phi_{\mu}\lr{\rr} },
	\end{multline}
	where the coordinate $x$ defined along the displacement may be expressed by $x_i$ denoting the coordinates along the [100], [010], and [001] crystallographic axes, $x = \sum_i c_i x_i$.
	At this point the reason for keeping $m$ inside the expression becomes evident, as, under the sum over valleys, we need to replace the operator composed of differentiation and inversed $m$ with
	\begin{equation}
		 \frac{1}{m}\frac{\partial}{\partial x} = \sum_i c_i \, \sum_\mu \delta_{\mu\nu} \, \frac{1}{m_{i,\nu}} \, \frac{\partial}{\partial x_i},
	\end{equation}
	where $\delta_{\mu\nu}$ is the Kronecker delta.
	In the calculation, we use the following values for the effective mass, $m_\perp = 0.191$ and $m_\parallel = 0.916$, where $m_{i,\mu} = m_\parallel$ applies if $\widehat{x}_i = \widehat{\mu}$, i.e., valley $\mu$ is oriented along $x_i$, and $m_{i,\mu} = m_\perp$ otherwise.
	
	For donor displacements in directions where valley interference leads to oscillatory behavior \cite{KoillerPRL2001}, we observe a significant variation of the results with respect to the location of the integration plane.
	To circumvent this problem, we introduce averaging over the position of $S$.
	This may be done in a mathematically elegant manner by transforming \eqref{gradtodx} into a volume integral with a Dirac delta constraint
	\begin{equation}\label{eq:dstod3r}
		\int_S \mathrm{d}S \, f\lr*{\rr} = \int_V \mathrm{d}^3\rr \, \delta\lr*{x-x_S}\, f\lr*{\rr},
	\end{equation}
	where $f\lr*{\rr}$ is the integrand, $x_S = d/2$ is the nominal position of $S$, $V$ is the volume of the system (calculation box), and then replacing $\delta\lr*{x}$ with its broadened representation with a finite support, like \cite{PazPSSB2006}
	\begin{align}\label{eq:deltabroad}
		{\delta}(x) &\simeq \frac{15}{16\sigma}\Lr*{1-\lr*{ \frac{x}{\sigma} }^2 }^2 \, \theta\lr*{ \sigma - \abs{x} } \nonumber\\
		&\equiv \widetilde{\delta}(x) \, \theta\lr*{ \sigma - \abs{x} }.
	\end{align}
	Here, $\sigma$ is the broadening and $\theta(x)$ is the Heaviside step function.
	Substituting \eqref{deltabroad} to \eqref{dstod3r}, we obtain
	\begin{multline}
		\int_V \mathrm{d}^3\rr \, f\lr*{\rr}\,\delta\lr*{x-x_S}
		\simeq \iint_{-\infty}^{\infty} \mathrm{d}y\mathrm{d}z \int_{-\sigma}^{\sigma} \mathrm{d}x \, f\lr*{\rr}\,\widetilde{\delta}\lr*{x-x_S},
	\end{multline}
	which is the integral we evaluate numerically in a finite box.
	Details are given in Appendix~\ref{sec:details}.
	While it is a triple integral, the smoothly decaying $\widetilde{\delta}$ constraint with $\sigma$ not exceeding a few lattice constants makes the increase of the numerical cost insignificant.
	The estimation of the computational cost of the calculation is described in Appendix~\ref{sec:calctime}.
	We find the time needed for satisfactory convergence of results to be comparable to the standard method.
	
\section{General results and discussion}\label{sec:results}
	In this section, we present the evaluated hopping integrals between pairs of identical (Sec.~\ref{sec:intra}) and different (Sec.~\ref{sec:inter}) orbitals on the two sites for a range of distances along relevant crystallographic directions, discuss in detail an exemplary case where the standard method encounters problems related to the central cell correction (Sec.~\ref{sec:pxpy}), and finally present and characterize calculated eigenstates of donor pairs (Sec.~\ref{sec:full}).

\subsection{Intraorbital hopping}\label{sec:intra}
	\begin{figure}[t]
		\includegraphics[width=1.0\linewidth]{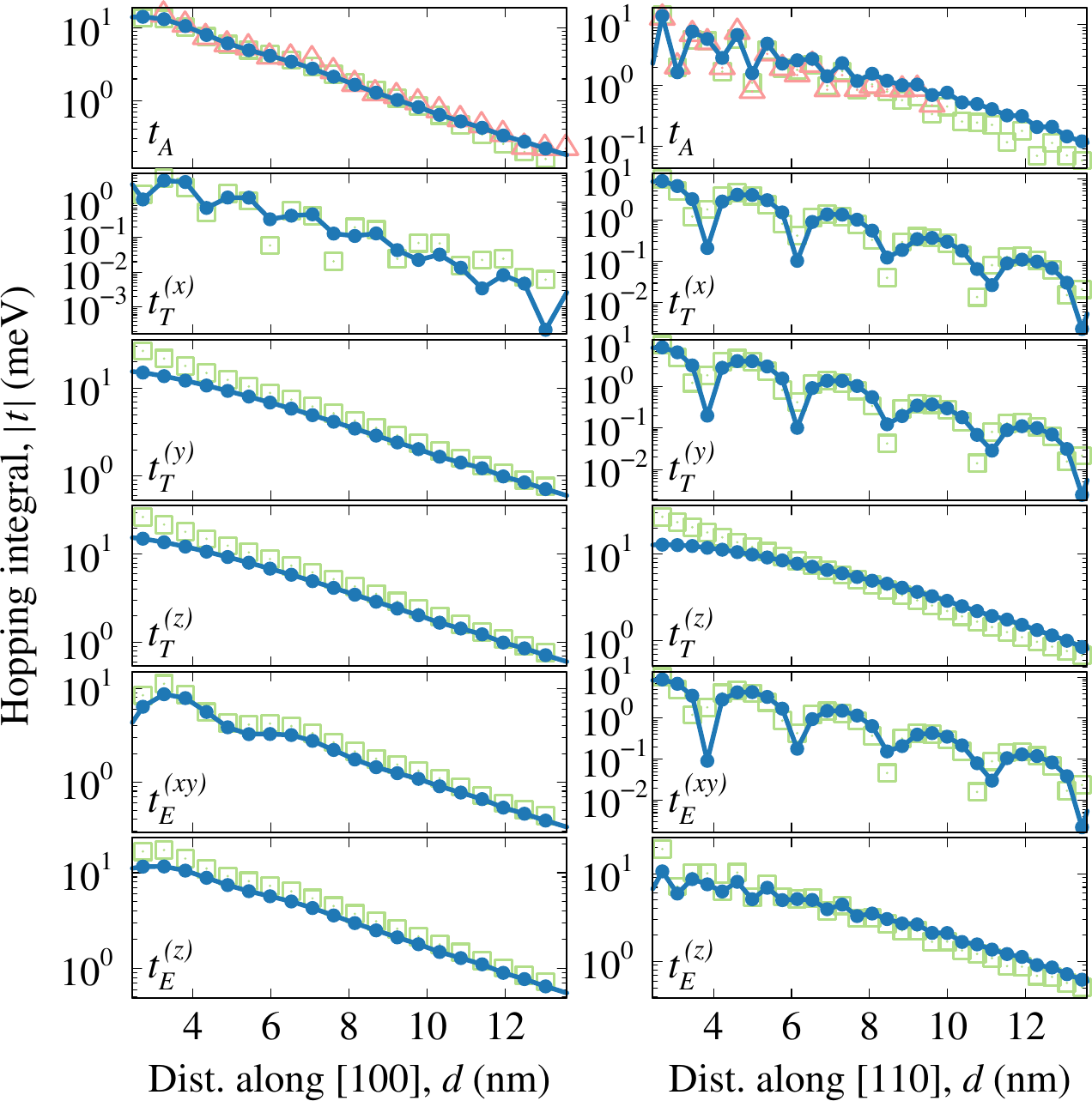} %
		\caption{\label{fig:hopsame}(Color online) Intraorbital hopping. Same-orbital hopping integrals (full symbols; {\color{cbblue}$\bullet$}) calculated as a function of donor distance along the [100] (left column) and [110] (right column) crystallographic directions. Each row of panels is for one of the six ground-state orbitals, as marked. For comparison, empty squares ({\color{cblgreen!90!black}$\bm{\square}$}) mark results obtained using the H\"uckel theory; in the first row, empty triangles ({\color{cblred!90!black}$\bm{\triangle}$}) show results from \oncite{LePRB2017a}. Lines are to guide the eye.} %
	\end{figure}

	We begin the presentation of results by showing in \figref{hopsame} the calculated hopping integrals between identical orbitals at the two donors.
	The values are shown as a function of the displacement: along [100] on the left, and along [110] on the right.
	To verify the accuracy of our calculation, we also plot with empty symbols the results of a standard computation employing the H\"uckel (tight-binding) theory \cite{AshcroftBook1976}. \newtext{We refer to this method as "standard" throughout the paper and describe it in Appendix~\ref{sec:huckel}.}
	In the top row, where hopping in the ground $A_1$ orbital is considered, we also show with triangles the available data from the literature \cite{LePRB2017a} (we acquired the values by digitizing the linear-scale plot from the original paper, which could lead to some degree of inaccuracy).
	We find the results to be in an overall very good agreement with these datasets.
	
	In the following panels, where tunneling between higher-energy orbitals is considered, we can compare only to the H\"uckel theory results obtained by us.
	Also here, the agreement is satisfactory.
	In particular the oscillations in hoppings along the [110] displacement are well reproduced.
	Apart from the irregular oscillation arising from valley interference and present in $t_A$ and $t_E^{(z)}$, we deal also with a regular one with longer period observed in $t_T^{(x)}$, $t_T^{(y)}$, and $t_E^{(xy)}$.
	It results from the in-plane excited-state character (i.e., presence of a node) of the wave function envelopes in these orbitals.
	The main discrepancies between the two methods are that hoppings calculated using Bardeen's theory are generally slightly lower at short distances while being qualitatively similar, and only the dependence of $t_{T}^{(z)}$ on distance is a bit different.

	At this point, we need to underline that both methods are approximate, and there is currently no experimental data we could compare to.
	\newtext{On the one hand, Bardeen's theory originally assumes a low-overlap system.
	On the other, it involves only the wave function and thus it is free of problems that arise when the potential is directly used.
	Moreover, in Sec.~\ref{sec:corr}, we derive a correction to Bardeen's formula that enhances its accuracy at short distances.
	The main discrepancy noticed above is then fixed.}
	In the H\"uckel tight-binding approach, one neglects the background potential and only considers those from the two sites.
	An additional complication arises for a donor system, as the exact form of the donor potential is unknown, and regular $1/r$ Coulomb potential augmented by central cell corrections is used.
	While this approach yields correct energy levels with wave functions having all expected properties, the physicality of phenomenologically introduced central cell corrections \cite{NingPRB1971,PantelidesPRB1974} is disputable.
	For this reason, it is excluded in the calculation of hopping integrals within the H\"uckel model \cite{LePRB2017a}.
	However, using the pure $1/r$ potential does not have to be exact either, as it has too high symmetry and implicitly assumes a uniform medium characterized by a dielectric constant.
	This approximation is not obvious on the length scales of a few lattice constants.
	As we show in the following in a specific example, using the pure $1/r$ potential in the standard method can lead to qualitatively incorrect results, while the inclusion of the central cell correction turns out to be numerically challenging.
	Thus, we treat the agreement of the results obtained within these two approaches as a mutual confirmation rather than a benchmark against a reference.
		
\subsection{Interorbital hopping}\label{sec:inter}
	\begin{figure}[tb]
		\includegraphics[width=1.0\linewidth]{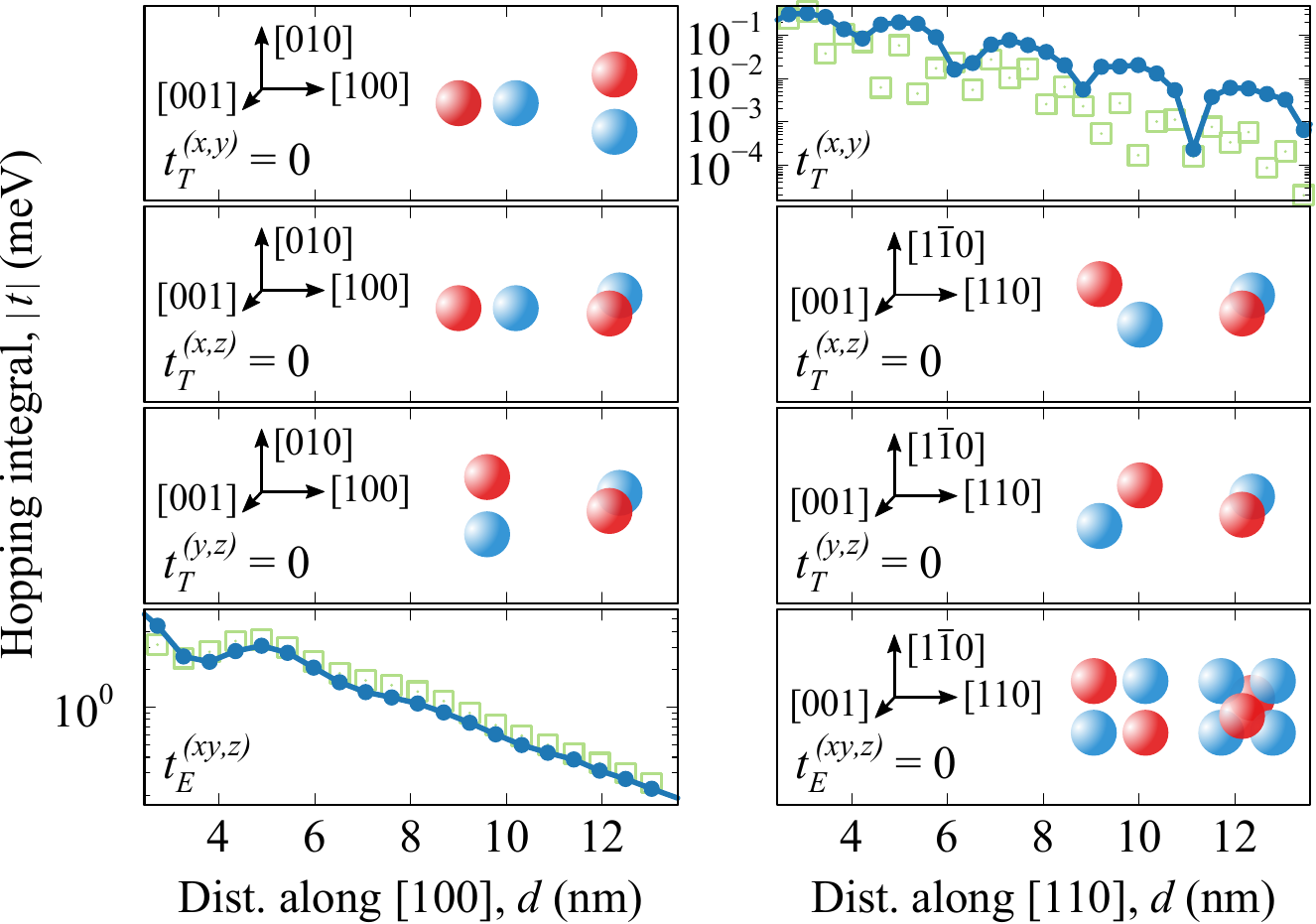} %
		\caption{\label{fig:hopdiffa}(Color online) Degenerate interorbital hopping. Hopping integrals between pairs of different degenerate orbitals within the $T_2$ and $E$ manifolds (full symbols; {\color{cbblue}$\bullet$}) calculated as a function of donor distance along the [100] (left column) and [110] (right column) crystallographic directions. Each row of panels is for a different orbital pair, as marked. For comparison, empty symbols ({\color{cblgreen!90!black}$\bm{\square}$}) mark results obtained using the H\"uckel theory. Lines are to guide the eye. For vanishing cases, the underlying symmetry of wave functions is schematically shown; the two colors mark the sign of wave-function envelopes.} %
	\end{figure}
	
	Having established this, we proceed to the evaluation of hopping between pairs of different orbitals.
	First, we consider degenerate pairs within the $T_2$ and $E$ manifolds.
	Here, most of the integrals vanish due to symmetry.
	In terms of \eqref{bardeen} and \eqref{gradtodx}, it happens when the product of the two orbitals is antisymmetric in the plane perpendicular to the displacement, thus if they differ in the in-plane parity.
	Note that the differentiation in \eqref{gradtodx} changes the parity only in the lateral direction, which is irrelevant in this regard.
	As we consider (001)-plane displacements, $t_{T}^{(x,z)}$ and $t_{T}^{(y,z)}$ vanish for any direction since the orbitals differ in the $z$-axis parity.
	On the other hand, $t_{T}^{(x,y)}$ may be nonzero for any displacement direction other than [100] and [010], for which it vanishes.
	The pair of orbitals from the $E$ manifold gives a nonzero hopping $t_{E}^{(xy,z)}$ for all directions except the diagonal ones: [110] and [1$\bar{1}$0].
	Thus, we calculate the two nonvanishing hopping integrals: $t_{T}^{(x,y)}$ for $\bm{d}\parallel$[110] and $t_{E}^{(xy,z)}$ for $\bm{d}$ along [100].
	The calculation is similar as previously, and the results are presented in \figref{hopdiffa}.
	Again, on the left, we show results for displacement along [100] and for [110] on the right, while each row is for a different orbital pair.
	In the vanishing cases, instead of plots, we show schematic diagrams visualizing the difference in parity that makes the coupling forbidden.
	For $t_{E}^{(xy,z)}$ along [100], we find good agreement with the standard method, including the non-monotonic behavior at short distances.
	Notably, the values are significant and comparable with same-orbital hopping.
	
	In the case of $t_{T}^{(x,y)}$ along [110] (top right panel in \figref{hopdiffa}), we face an issue.
	To obtain nonvanishing hopping values using the standard method, we need to explicitly take into account the central cell correction in the on-site potential.
	It applies to the entire range of distances, including those much larger than the spatial extent of the correction potential.
	As the latter is very local compared to the integration domain, the need for its inclusion creates a great computational challenge.
	We find the computed values to be significantly sensitive to the integration volume and other computational details.
	Thus, we cannot confirm that the obtained strongly oscillating result is quantitatively correct.  
	On the other hand, using Bardeen's theory, we get a well-converged result without any additional treatment.
	The integral is considerably small compared to others, but it exhibits a meaningful slow oscillation similar to those observed above in same-orbital couplings for $p$-type orbital envelopes.
	We discuss the calculation of this integral in more detail in the following subsection.

	\begin{figure}[tb]
		\includegraphics[width=1.0\linewidth]{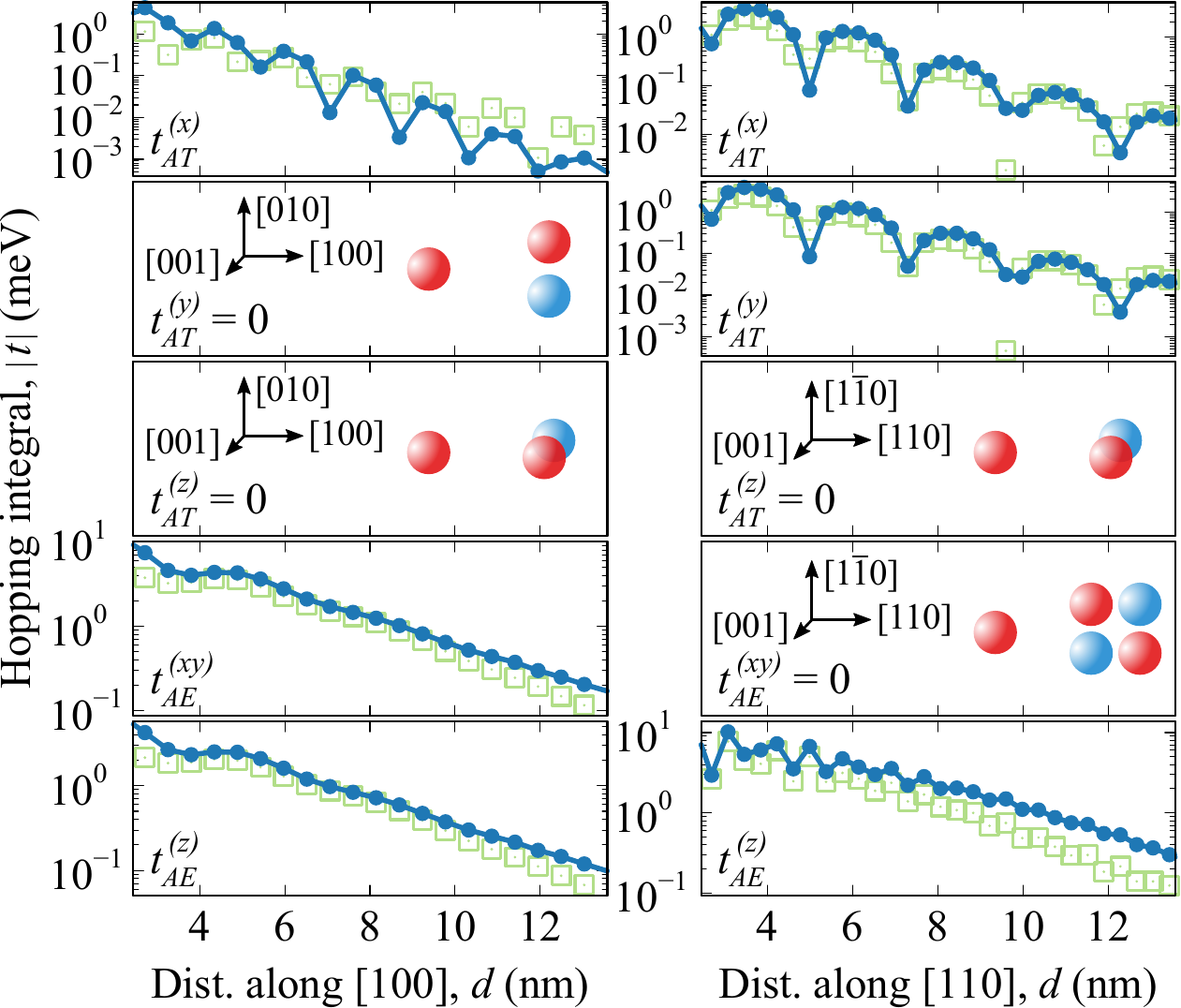} %
		\caption{\label{fig:hopdiffb}(Color online) Nondegenerate interorbital hopping. As in \figref{hopdiffa}, but for hopping integrals between pairs of orbitals from different manifolds: from $A_1$ to $T_2$ and $E$.} %
	\end{figure}

	\begin{figure}[tb]
		\includegraphics[width=1.0\linewidth]{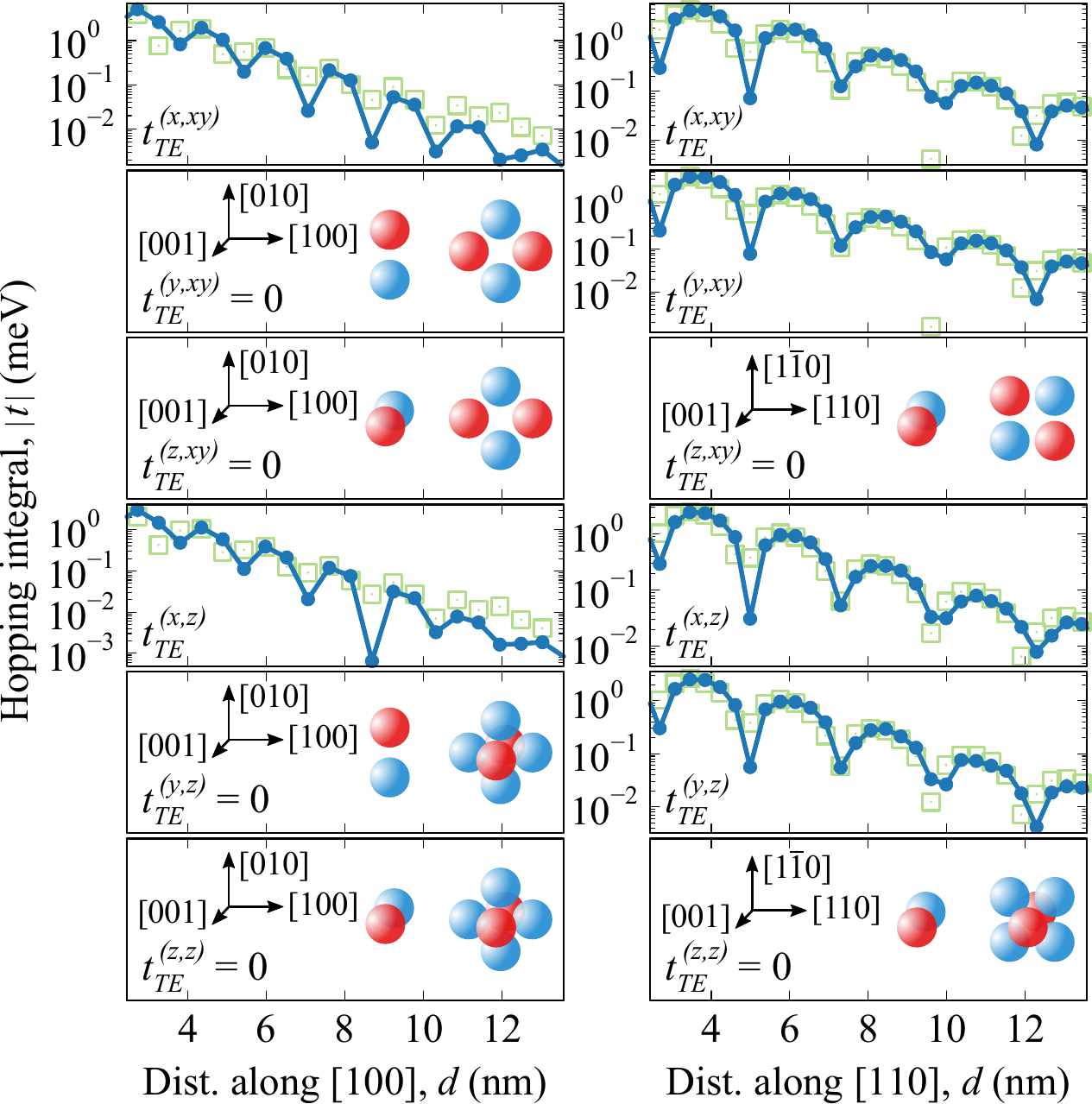} %
		\caption{\label{fig:hopdiffc}(Color online) Nondegenerate interorbital hopping. As in \figref{hopdiffa}, but for hopping integrals between pairs of orbitals from different manifolds: from $T_2$ to $E$.} %
	\end{figure}

	We are left with the calculation of hopping integrals connecting different manifolds.
	Also in this case, a number of couplings vanish due to symmetry.
	For those that may be nonzero, as nondegenerate orbital pairs are considered, we also need to evaluate the correction given in the second term of \eqref{bardeen}.
	Technically, for the problem in question, it is a half of the wave-function overlap multiplied by the energy splitting.
	The results, presented in \figref{hopdiffb} and \figref{hopdiffc} for two displacement directions as previously, are in similarly overall good agreement with the standard method, as it was for the same-orbital and degenerate interorbital hoppings except for $t_{T}^{(x,y)}$ along [110].
	The main difference between the results that may be found in \figref{hopdiffb} is the large-distance behavior of couplings between $T_2$ and $E$ orbitals: the standard method predicts a faster decay.
	A weaker opposite difference is also noticeable in $t_{AT}^{(x)}$ along [100].
	In \figref{hopdiffc}, only a minor shift in the oscillation phase is present in some of the hoppings. 
	Again, we notice that all nonvanishing hopping integrals are comparable not only with the same-orbital couplings but, at short distances, also with the orbital splitting.
	This explains the observed transition to the strong coupling regime below $\sim6$~nm \cite{KlymenkoJPCM2014,SaraivaJPCM2015,GamblePRB2015}.

\subsection{Calculation of $t_{T}^{(x,y)}$ along [110]}\label{sec:pxpy}
	\begin{figure}[tb]
		\includegraphics[width=0.75\linewidth]{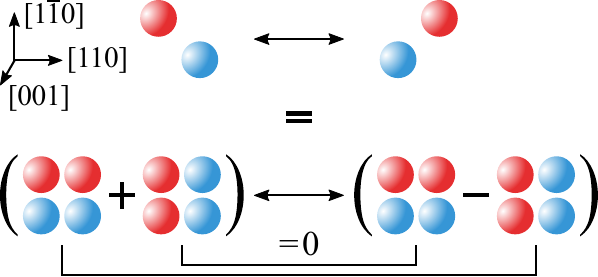} %
		\caption{\label{fig:pxpy_scheme}(Color online) Schematic presentation of the expansion of $t_{T}^{(x,y)}$ into even and odd contributions with respect to the plane normal to $\bm{d}\parallel$[110]. Vanishing contributions are marked.} %
	\end{figure}

	\begin{figure}[tb]
		\includegraphics[width=0.9\linewidth]{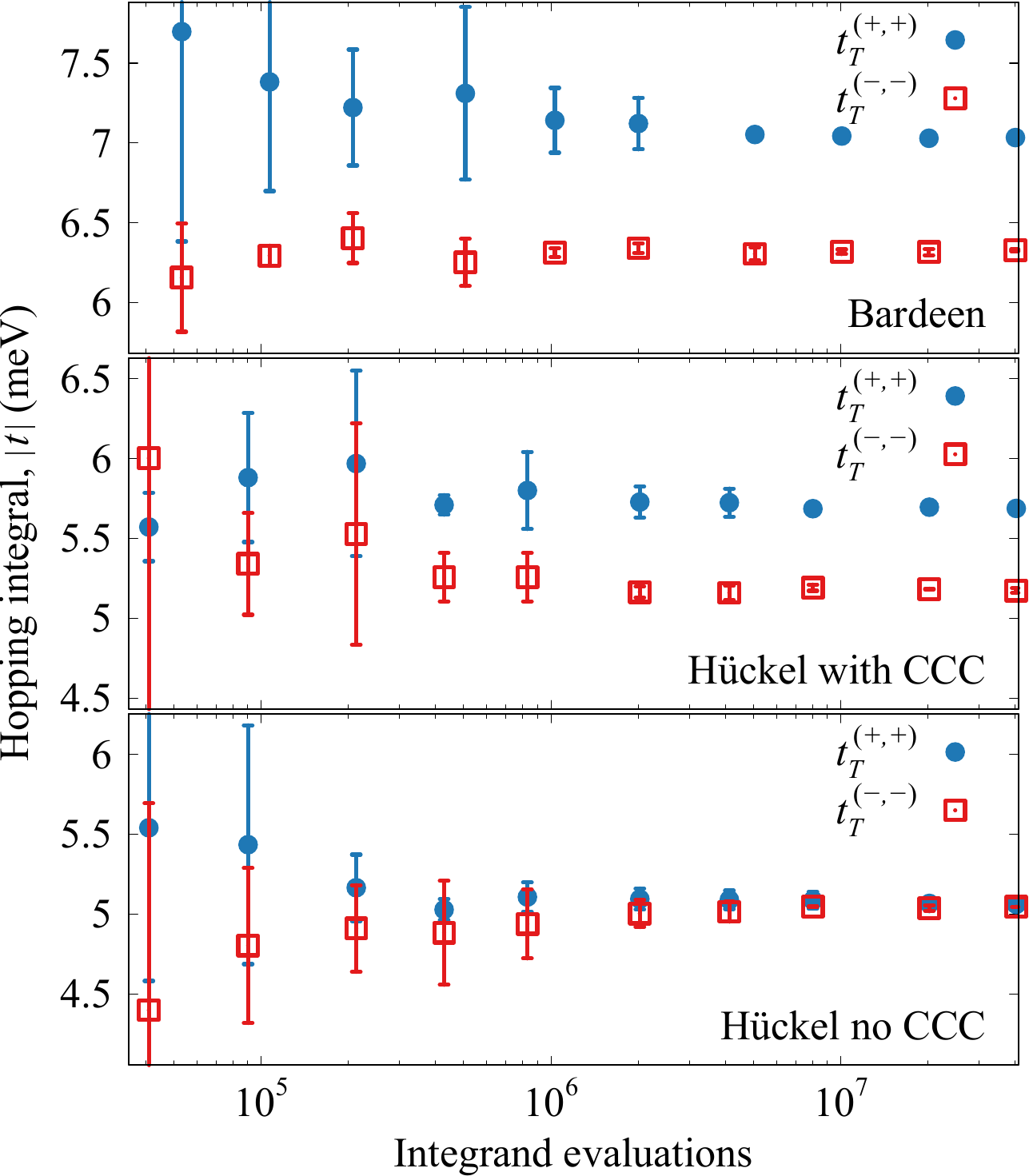} %
		\caption{\label{fig:pxpy110}(Color online) Evaluation of $t_{T}^{(x,y)}$ for separation along [110]. Hopping integrals $t_{T}^{(+,+)}$ and $t_{T}^{(-,-)}$ for a fixed donor distance along the [110] crystallographic direction as a function of the number of integrand evaluations (integration precision). The top panel shows the results obtained using Bardeen's theory; the middle and bottom panels are for the H\"uckel method with and without the central cell correction (CCC), respectively.} %
	\end{figure}

	Let us go back to the calculation of $t_{T}^{(x,y)}$ along [110], where an issue arises.
	Trying to converge the calculation using the standard method, we obtain results that seem to tend to zero in a weak manner (weakly decreasing values comparable to their uncertainty, behavior different than for hoppings vanishing due to symmetry).
	As we use Monte Carlo integration, and the integrand is highly oscillatory, this could mean that the integral vanishes.
	On the other hand, using Bardeen's theory, we get well-converged values, which are small compared to other integrals, but certainly do not vanish.
	As both methods are approximate, we cannot decide readily which of the results is correct.
	
	There is no evident symmetry-based argument for the vanishing of the given hopping.
	To get more insight, we define the mixed states
	\begin{equation}\label{eq:tpm}
		\ket[\big]{\,T_2^{(\pm)}} = \frac{1}{\sqrt{2}}\lr*{ \ket[\big]{\,T_2^{(x)}} \pm \ket[\big]{\,T_2^{(y)}} }
	\end{equation}
	with well-defined parity in the separation plane normal to $\bm{d}\parallel$[110], even and odd, respectively.
	Using the inverse transformation,
	\begin{equation}\label{eq:txy}
		\ket[\big]{\,T_2^{(x/y)}} = \frac{1}{\sqrt{2}}\lr*{ \ket[\big]{\,T_2^{(+)}} \pm \ket[\big]{\,T_2^{(-)}} },
	\end{equation}
	we may rewrite the hopping integral in question as
	\begin{align}\label{eq:txypm}
		t_{T}^{(x,y)} &= \frac12 \lr*{ t_{T}^{(+,+)} - \,t_{T}^{(-,-)} - \,t_{T}^{(+,-)} + \,t_{T}^{(-,+)} } \nonumber\\
		&= \frac12 \lr*{ t_{T}^{(+,+)} - \,t_{T}^{(-,-)} },
	\end{align}
	where the last two terms, $t_{T}^{(+,-)}$ and $t_{T}^{(-,+)}$, vanish as they couple states of different parity.
	Thus, the transformation allows us to explicitly remove two vanishing contributions, and express $t_{T}^{(x,y)}$ as a difference of two definitely finite integrals, as schematically shown in \figref{pxpy_scheme}.
	We expect the integral to be small, as it is given by a difference of two similar terms. However, there is no reason why it should vanish.
	
	In \figref{pxpy110}, we plot the two contributions, $t_{T}^{(+,+)}$ and $t_{T}^{(-,-)}$, as a function of the number of integrand evaluations, i.e., we show how they converge.
	The bottom panel shows the results obtained for the standard method with the $1/r$ potential used.
	The two contributions tend to have the same value, and hence $t_{T}^{(x,y)}$, given by their difference, vanishes.
	On the other hand, Bardeen's theory gives us a finite difference and thus nonvanishing hopping integral, as shown in the top panel.
	Looking for the reason for this discrepancy, we focus on the differences between the two methods.
	In both cases, we use the same wave functions.
	In Bardeen's theory, they are the only ingredient for the calculation, while in the standard H\"uckel method, the on-site potential is integrated between these functions.
	Here, a subtle difference arises in the symmetry of the problem in the two methods.
	The wave functions implicitly inherit the tetrahedral symmetry of the full donor potential (including the central cell correction), while the symmetry of the $1/r$ potential is higher.
	To verify if this is the source of the problem, we repeat the H\"uckel-method calculation, this time adding the central cell correction to the integrated potential.
	The result is shown in the middle panel, where the two contributions indeed show a finite difference, as those calculated using Bardeen's theory.
	We need to emphasize that, in this specific case, the central cell correction influences the results in the entire range of donor distances.
	The lack of this symmetry-breaking correction changes the result qualitatively, giving an artificially vanishing hopping.

	In general, the solution should be straightforward.
	One needs just to take into account the central cell correction.
	However, its inclusion in numerical integration is challenging.
	The explicit form of the correction potential for a donor is \cite{GamblePRB2015}
	\begin{equation}
		U_{\mathrm{cc}}\lr{\rr} = A_0 e^{-r^2/(2a^2)}+A_1\sum_{i=1}^4 e^{-\abs{\bm{r}-b\bm{t}_i}^2/(2c^2)},
	\end{equation}
	where $\bm{t}_i\in\LR{ (1,1,1),(-1,1,-1),(1,-1,-1),(-1,-1,1)}$ are the tetrahedral directions of the bonds, $A_0=-1.2837$~meV, $A_1=-2642.0$~meV are the amplitudes, $a=0.12857$~nm is the spatial extent of the symmetric part, while $b=0.21163$~nm, and $c=0.09467$~nm are the displacement and extent of the non-spherical parts.
	The last parameter shows that symmetry-breaking contributions are very local, as their spatial extension is tiny compared to the volume of the two-donor system, hence also the integration domain.
	Thus, obtaining quantitatively correct results when they critically depend on this potential is at least challenging.
	We observe a substantial variation of calculated values with the integration domain size (in the range where it should be already large enough) as well as with the specific integration algorithm used.
	Because of this, we are unable to confirm the quantitative accuracy of these specific H\"uckel-method results.
	
	Thus, while in general correct results can be obtained in the standard method if one is careful about the potential used, it may occur to be computationally unfeasible. 
	In contrast, an advantage of Bardeen's theory is revealed here, resulting from its dependence on the barrier-region parts of wave functions only.

\subsection{Donor pair eigenstates}\label{sec:full}
	\begin{figure}[tb]
		\includegraphics[width=1.0\linewidth]{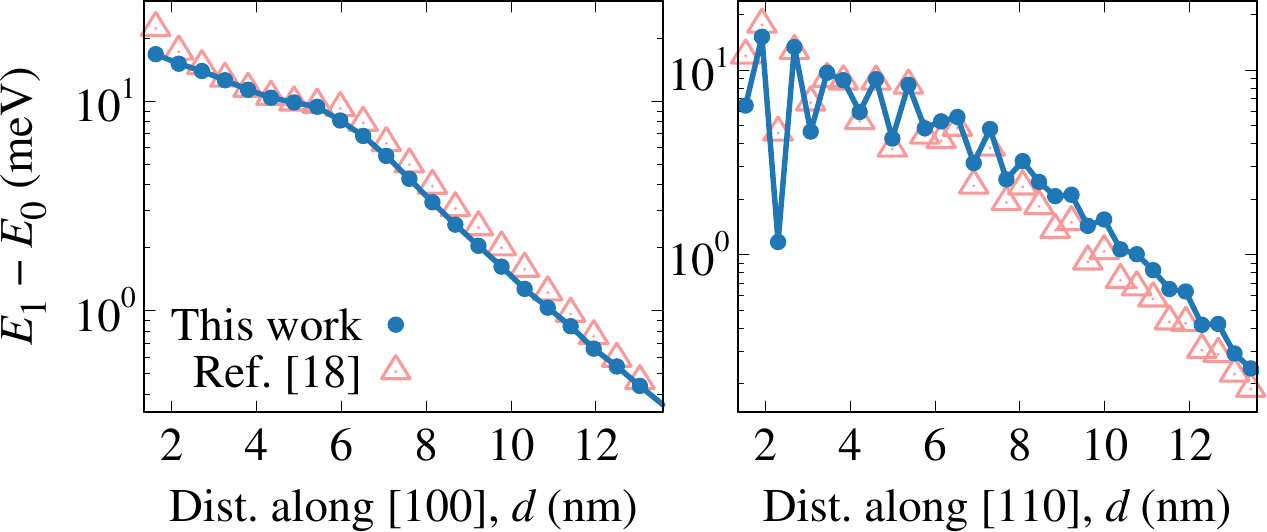} %
		\caption{\label{fig:hopfull}(Color online) Donor pair excited-ground state splitting. Energy splitting between two lowest-energy eigenstates of a donor pair (full symbols; {\color{cbblue}$\bullet$}) calculated as a function of donor distance along the [100] (left column) and [110] (right column) crystallographic directions. For comparison, empty triangles ({\color{cblred!90!black}$\bm{\triangle}$}) show the result of a full calculation from \oncite{GamblePRB2015}. Lines are to guide the eye only.} %
	\end{figure}
	
	Having all the hopping integrals calculated and knowing the single-donor orbital energies, we may finally construct the total Hamiltonian from \eqref{hamiltonian}.
	By diagonalizing it, we obtain the energy spectrum of a donor pair.
	In \figref{hopfull}, we show the energy splitting between the ground and first excited states calculated as a function of donor distance in the [100] and [110] directions.
	In this case, we may benchmark our results against a full two-donor EMT calculation from \oncite{GamblePRB2015} shown with empty triangles.
	We find the result to be in a very good qualitative agreement, with some minor quantitative differences, mainly in the medium distance regime.
	The transition to the strong coupling regime, visible for the [100] displacement as a kink at $d\simeq \SI{6}{\nano\metre}$, is reproduced correctly.
	Also, the oscillations for [110] displacement are in phase for the entire distance range.
	This agreement confirms the suitability of the proposed method to calculate the eigenstates of pairs and clusters of admixtures based only on the wave functions of a single donor.

	\begin{figure}[tb]
		\includegraphics[width=1.0\linewidth]{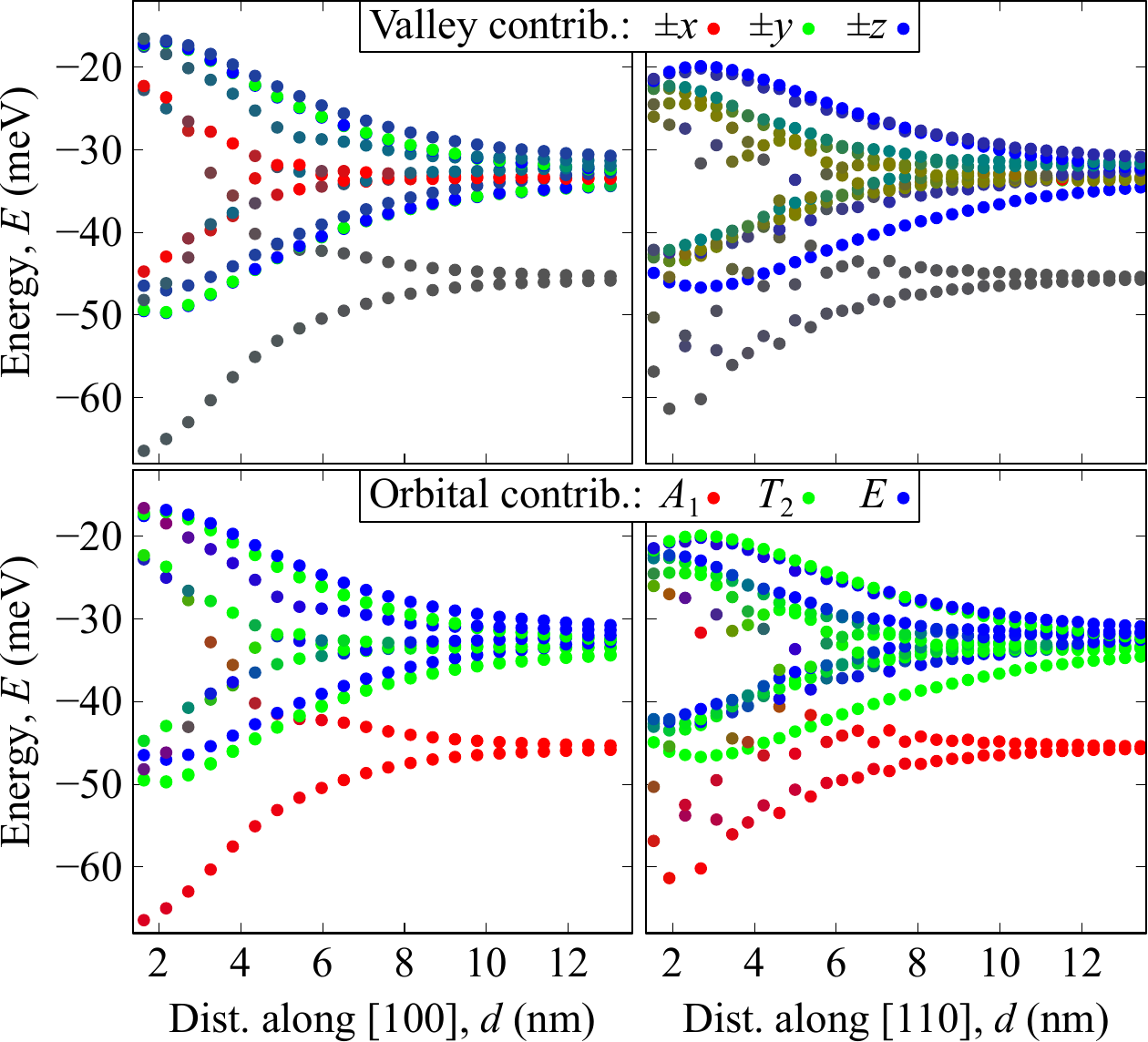} %
		\caption{\label{fig:fulleig}(Color online) Donor pair energy spectra. Energy (with respect to the Si conduction band edge) of the twelve lowest eigenstates of a donor pair calculated as a function of donor distance along the [100] (left column) and [110] (right column) crystallographic directions. The red, green, and blue components of the point color (additive RGB model) show the contributions of $\pm x$, $\pm y$, and $\pm z$ valleys (top panels) or $A_1$, $T_2$, and $E$ orbitals (bottom), respectively. The apparent random ordering of blue and green points in the top left panel is due to the degeneracy of $T_2^{(y)}$ and $T_2^{(z)}$ levels.} %
	\end{figure}
		
	Next, in \figref{fulleig}, we plot entire calculated spectra, i.e., twelve donor-pair eigenstates, again as a function of [100] and [110] donor separation.
	To characterize the eigenstates, we color-code the information on their composition: in the top row of panels, valley contributions are shown with the red, green, and blue color components, while in the bottom row, we similarly present the contribution of orbital types.
	When the tunnel coupling is relatively weak, i.e., at larger distances $d>$~nm, there is no evident orbital mixing, and each of the orbitals forms its own bonding and antibonding eigenstates.
	The two lowest-energy levels are then such states composed mainly of the lowest $A_1$ orbitals.
	Their splitting is equal to twice a quantity, which may be considered as an effective ground-state tunnel coupling $t_{\mr{eff}}$.
	For strong coupling, when hopping integrals are comparable or greater than orbital splittings, the ordering of states is affected, and the antibonding state with dominant $A_1$ contribution is no longer the first excited state \cite{KlymenkoJPCM2014,SaraivaJPCM2015}.
	This transition is reflected in the kink seen in \figref{hopfull}.
	Additionally, the antibonding state mixes considerably with other orbitals.
	Consequently, determining the proper value of effective hopping $t_{\mr{eff}}$ for $d<6$~nm requires more detailed research. Moreover, the very issue of the applicability of single-band models for closely spaced dopant arrays also requires consideration.
	These issues will be addressed in our upcoming work.
	
\newtext{
\section{Applicability of the method to overlapping potentials}\label{sec:corr}
	In this section, we rederive the formula for the hopping without assuming non-overlapping potentials, to assure its applicability to long-range potentials as those studied here.
	Additionally, we show that \eqref{bardeen}{} can be corrected to better describe the hopping at short distances.
	
	Bardeen's theory aims to calculate the tunneling current (or tunneling transition rate) between the initial and final states being the eigenstates of two potentials.
	The result for the rate is perturbative and has the form of Fermi's golden rule, which can also be obtained via standard time-dependent perturbation theory \cite{ReittuAJoP1995}.
	In this aspect, one deals with standard limitations: the matrix element has to be a small perturbation, and the calculated rate is valid at long enough time scales.
	
	Here, we do not study time-dependent phenomena, as we are only interested in the tunneling Hamiltonian.
	Thus, we exploit only a part of Bardeen's derivation showing that the matrix element can be calculated as a surface integral of the transition probability current $J_{ij}\mkern-1mu\lr{\rr}$.
	For this, the conditions of the perturbation theory do not apply.
	
	An additional assumption is made in Bardeen's theory that the two potentials do not overlap.
	For long-range potentials like $1/r$ considered here, this is not fulfilled.
	Here, we rederive the formula without this assumption and show that the surface integral from \eqref{bardeen} is, in fact, generally valid for the calculation of tunnel coupling.
	For non-overlapping potentials, it additionally conforms to the transfer Hamiltonian matrix element as defined by Bardeen, which allows then for calculation of the transition rate.
	
	Let $H_1 = T + U_1$ and $H_2 = T + U_2$ be the Hamiltonians of the two isolated parts of the system, where $T = -\hbar^2\nabla^2/2m$ is the kinetic energy, and $U_i$ is the $i$th potential.
	$H_1$ differs from $H_2$ by the position at which its potential is centered.
	Their ground states are $H_i\psi_i=E\psi_i$.
	Let us now assume that we are in the range of distances at which the hopping integral is well defined.
	This means that the lowest-energy eigenstates of $H=T+U_1+U_2$, i.e., for the pair of potentials (sites), are given by the bonding and antibonding superpositions of single-site ground states
	\begin{equation}\label{eq:app-hopp-def}
		H \, \frac{\psi_1\pm\psi_2}{\sqrt{2}} = \lr{ E\pm t } \, \frac{\psi_1\pm\psi_2}{\sqrt{2}},
	\end{equation}
	which are split by twice the hopping integral $t$.
	From this, by adding/subtracting by sides, we get
	\begin{subequations}\label{eq:app-hopp-eqns}
		\begin{align}
			H\psi_1 = E\psi_1 + t\psi_2, \\
			H\psi_2 = E\psi_2 + t\psi_1.
		\end{align}
	\end{subequations}
	Now, we left-multiply the first equation by $-\psi_2^{*}$, conjugate the second one and multiply it by $\psi_1$, add equations by sides and integrate over half-space $x>x_0$.
	This yields
	\begin{align}\label{eq:app-hopp-diff}
		\int_{x_0}^\infty \!\!\mathrm{d}x \, \psi_1\lr{x} H \psi_2^{*}\lr{x} &{} - \int_{x_0}^\infty \!\!\mathrm{d}x \, \psi_2^{*}\lr{x} H \psi_1\lr{x}  \nonumber \\
		= \int_{x_0}^\infty \!\!\mathrm{d}x &{}\Big[  t\abs*{\psi_1\lr{x}}^2 - t\abs*{\psi_2\lr{x}}^2   \\
		&{}+ E\psi_1\lr{x}\psi_2^{*}\lr{x} - E\psi_2^{*}\lr{x}\psi_1\lr{x} \Big],\nonumber
	\end{align}
	where terms in the last line are identical and cancel out.
	On the left-hand side, potential terms from the two integrals also cancel out, and, for the kinetic-energy part, by integrating one of the terms by parts twice, we get
	\begin{equation}
		\int_{x_0}^\infty \!\!\mathrm{d}x \, \psi_1\lr{x} T \psi_2^{*}\lr{x} - \int_{x_0}^\infty \!\!\mathrm{d}x \, \psi_2^{*}\lr{x} T \psi_1\lr{x} = J_{ij}\mkern-1mu\lr{x_0},
	\end{equation}
	where $J_{ij}$ is the transition probability current density from Bardeen's theory, given in \eqref{trans-prob-curr}. 
	Finally, we get
	\begin{align}\label{eq:bardeen-corr}
		t &{}= - J_{ij}\mkern-1mu\lr{x_0} \LR*{ \int_{x_0}^\infty \mathrm{d}x \Lr*{ \abs*{\psi_2\lr{x}}^2 - \abs*{\psi_1\lr{x}}^2 } }^{-1} \nonumber\\
		&{}\equiv - J_{ij}\mkern-1mu\lr{x_0} \, R^{-1},
	\end{align}
	where the minus sign comes from the fact we treated $t$ as positive, and we defined $R = 1-\rho_1-\rho_2$, with $\rho_i$ being the tails of the two probability densities on the sides of the division point opposite to the location of the given site.
	Thus, the hopping integral is given by the matrix element from Bardeen's theory up to a multiplicative factor that tends towards unity for a low-overlap system.
}
	
	\newtext{We can repeat this reasoning for the case of no degeneracy, in which the eigenstates are unequal superpositions of the initial states.
	Assuming the orbital splitting $\Delta E=E_2-E_1$ to be small and thus keeping only linear terms in $\Delta E$, we write 
	\begin{align}\label{eq:app-hopp-def2}
		H \, \lr*{\alpha\psi_1+\beta\psi_2} &{}= \phantom{-} t \, \lr*{\alpha\psi_1+\beta\psi_2}, \nonumber \\
		H \, \lr*{-\beta\psi_1+\alpha\psi_2} &{}= -t \, \lr*{-\beta\psi_1+\alpha\psi_2},
	\end{align}
	where $\alpha = \cos\lr{\theta/2}$, $\beta = \sin\lr{\theta/2}$ are the superposition coefficients (real for real $t$) with the mixing angle $\theta=\mr{atan}\lr{2t/\Delta E}$, and we put the mean energy to zero.
	By multiplying the first equation by $\alpha$ and the second one by $\beta$ and subtracting by sides (and vice versa, followed by adding), we get
	\begin{subequations}\label{eq:app-hopp-eqns2}
		\begin{align}
			H\psi_1 = \phantom{-}t\,\lr*{\alpha^2-\beta^2}\,\psi_1 + 2\alpha\beta\,t\,\psi_2, \\
			H\psi_2 =           -t\,\lr*{\alpha^2-\beta^2}\,\psi_2 + 2\alpha\beta\,t\,\psi_1.
		\end{align}
	\end{subequations}	
	As previously, we multiply the first equation by $-\psi_2^{*}$, conjugate the second one and multiply it by $\psi_1$, add by sides and integrate over half-space $x>x_0$, which gives
	\begin{equation}
		\!\!\!J_{ij}\lr*{x_0} = - t\left\lbrack 2\,\alpha\beta \, R +2\lr*{\alpha^2-\beta^2}\int_{x_0}^{\infty} \!\!\mathrm{d}x \, \psi_2^{*}\lr{x}\psi_1\lr{x} \right\rbrack.\!\!
	\end{equation}
	Keeping up to linear terms in $\Delta E$, we have $\alpha(\beta) \simeq 1/\sqrt{2} \mp \Delta E /4\sqrt{2}t$, and thus $\alpha\beta\simeq 1/2$ and $\alpha^2-\beta^2\simeq-\Delta E/2t$.
	With this, we arrive at the result for the hopping
	\begin{equation}\label{eq:bardeen-nondeg-corr}
		t = - \Lr*{ J_{ij}\mkern-1mu\lr{x_0} - \lr*{E_1-E_2}  \int_{x_0}^{\infty} \!\!\mathrm{d}x \, \psi_2^{*}\lr{x}\psi_1\lr{x} }\,R^{-1},
	\end{equation}
	where the second term on the right-hand side reproduces the correction for nondegeneracy from \eqref{bardeen}.
	The generalization of the above derivations to three dimensions is straightforward.
	It yields an integral of $J_{ij}\mkern-1mu\lr{\rr}$ over the division surface $S$ and a volume integral over the $V_{+}$ region in the second term, as in \eqref{bardeen}, and in the expression for $R$.
}
	
	\newtext{In this way, we have reproduced Bardeen's result without using the assumption of non-overlapping potentials.
	The overlap of wave functions is still not treated strictly, but we have obtained a correction accounting for it in the form of the $R$ factor.
	The latter depends on wave functions only, so the augmented method remains potential-free in the sense that one does not need to use donor/site potentials in the calculation explicitly.}
	
	\begin{figure}[tb]
		\includegraphics[width=\linewidth]{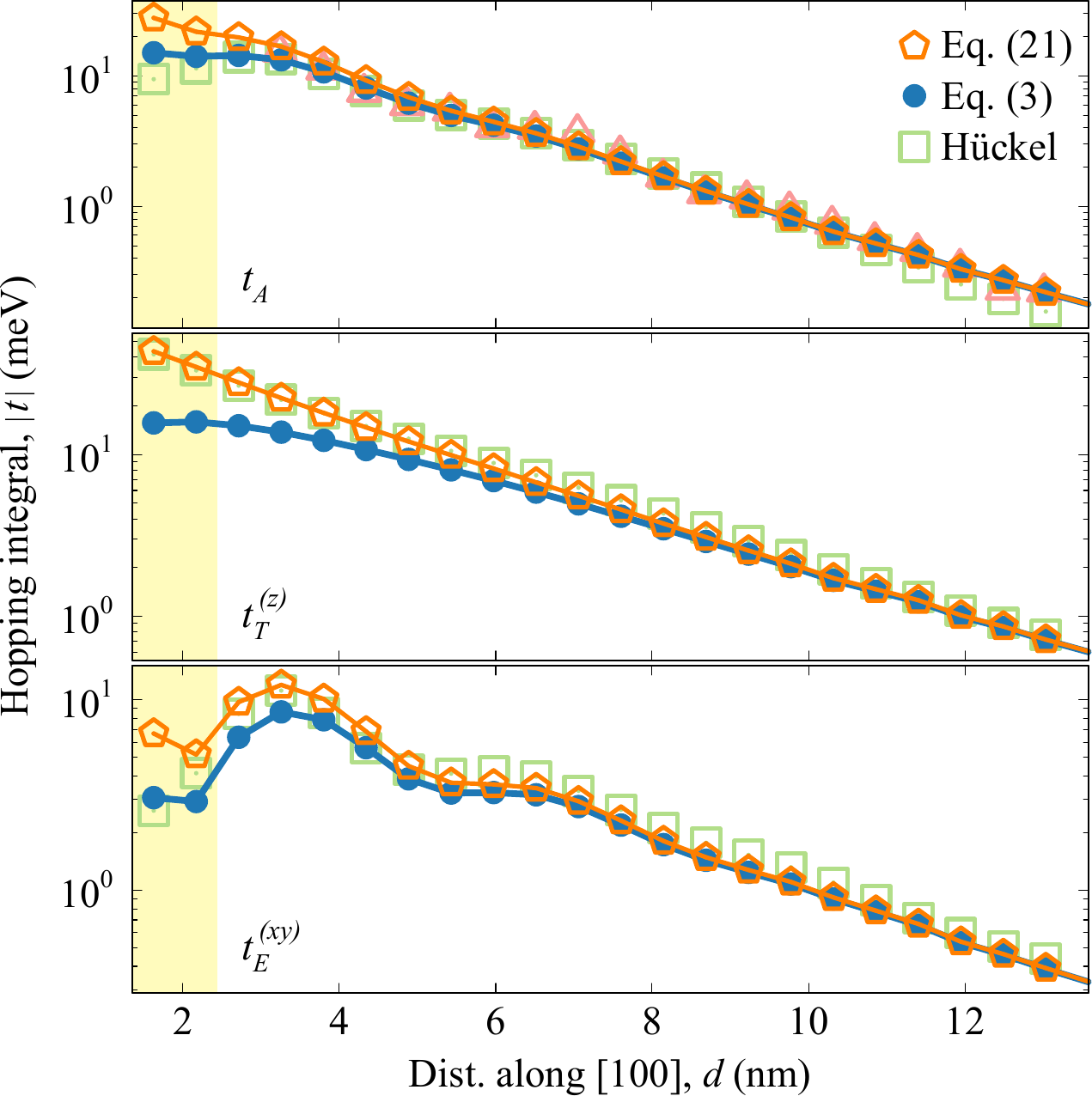} %
		\caption{\label{fig:overlap-corr}(Color online) Results corrected for the overlap. Selected same-orbital [100]-axis hopping integrals calculated with (empty pentagons; {\color{cborange}$\boldsymbol{\pentagon}$}) and without (full circles; {\color{cbblue}$\bullet$}) the $R$ factor from \eqref{bardeen-corr}. For comparison, empty squares ({\color{cblgreen!90!black}$\bm{\square}$}) mark results obtained using the H\"uckel theory; in the first row, empty triangles ({\color{cblred!90!black}$\bm{\triangle}$}) show results from \oncite{LePRB2017a}. The shaded area shows the range of distances not presented in previous plots. Lines are to guide the eye.} %
	\end{figure}	
	\newtext{In \figref{overlap-corr}, we show for selected cases that introducing this correction (empty pentagons) brings our results into even better agreement with H\"uckel tight-binding (empty squares), however, the difference it introduces is not significant overall down to distances of $d\simeq4$~nm.
	At around $d\simeq2.5$~nm our corrected results start to diverge from the ones from the H\"uckel method.
	This may be due to the lack of central cell correction in our H\"uckel-method calculation, but it could also be a sign of reaching the limit of applicability of our method.}
	
\section{Conclusions}\label{sec:conclusions}
	We have shown that hopping integrals (tunnel couplings) between phosphorus donors in silicon can be calculated with satisfactory accuracy using Bardeen's transfer method when orbital wave functions are known.
	\newtext{We have calculated both inter- and intraorbital tunnel matrix elements, which are essential for constructing multi-orbital lattice models of donor arrays.}
	We have also used these hoppings to form and diagonalize the two-donor Hamiltonian.
	With this, we have obtained the ladder of eigenstates and characterized their orbital and valley composition.
	
	In contrast to the commonly used H\"uckel theory, the method used by us does not involve integration with donor or lattice potentials.
	Instead, the matrix element is evaluated purely from the barrier-region parts of wave functions of the two states in question.
	\newtext{This turns out to be practically advantageous}, as we show that neglecting the central cell correction in the standard method may lead to qualitatively incorrect results, while its inclusion in the integration is computationally troublesome.
	\newtext{In contrast, wave functions obtained with the correction do not cause such problems in the proposed method.}
	
	\newtext{As the original derivation for the matrix element in Bardeen's theory exploits the assumption of non-overlapping potentials, which is not fulfilled for $1/r$ ones, we present a derivation that does not rely on this assumption.
	Additionally, we find a correction to the original expression, which extends the applicability of the method to shorter distances (higher wave function overlaps).}
	
	Using the available wave functions for the six orbitals forming the ground-state manifold in a Si:P donor, we have calculated tunneling matrix elements both for matching and different pairs of orbitals.
	The results are close to those obtained in a standard way, and, where available, we have compared them with data from the literature.
	While for crystallographic directions, where valley interference occurs, additional averaging is needed, the presented method has turned out to be of comparable computational cost to the standard one.
	Concerning this and the conceptual advantages it offers, we find the method to be competitive to the commonly used H\"uckel theory.
	Our work may serve as a benchmark of the method with a positive outcome and indicates its suitability for evaluating hopping integrals for lattice models when orbital wave functions are known or postulated.  

\acknowledgments
We acknowledge support from the National Science Centre (Poland) under Grant No. 2015/18/E/ST3/00583.

\appendix
\newtext{
\section{H\"uckel tight-binding}\label{sec:huckel}
	The standard method to which we compare in this study is based on H\"uckel's tight binding theory.
	In this approach, for a pair of orbitals, one on each of the donors, one solves a generalized eigenproblem
	\begin{equation}
		\begin{bmatrix}
			\alpha & \beta \\
			\beta^* & \alpha'
		\end{bmatrix}
		\begin{bmatrix}
			c_{a} \\
			c_{b}
		\end{bmatrix}
		= E
		\begin{bmatrix}
			1 & S \\
			S^* & 1
		\end{bmatrix}
		\begin{bmatrix}
			c_{a} \\
			c_{b}
		\end{bmatrix},
	\end{equation}
	where the diagonal and off-diagonal elements are given by $\alpha = \matrixel{1}{H}{1}$, $\alpha' = \matrixel{2}{H}{2}$, $\beta = \matrixel{2}{H}{1}$, with $\braket{\rr}{1/2}=\psi_{1/2}\lr{\rr}$, while $S = \braket{1}{2}$ is the overlap of the two states which are not fully orthogonal, and $H=T+U_1\lr{\rr}+U_2\lr{\rr}$ is the Hamiltonian of the two-site system with $T$ being the kinetic energy, and
	\begin{equation}
		U_{i}\lr{\rr} = - \frac{e^2} { 4 \pi \epsilon_0 \epsilon_{\mr{Si}}} \frac{1}{\abs{\rr-\rr_i}} + U_{\mathrm{cc}}\lr{\rr-\rr_i}
	\end{equation}
	the potential of the $i$th donor (we use the central-cell correction only in Sec.~\ref{sec:pxpy}).
	Note that $\ket{1}$ is the eigenstate of $H_1=T+U_1$, and $\ket{2}$ of $H_2=T+U_2$.
	Upon diagonalization, one arrives at the eigenvalues, and their difference determines the hopping.
	Conveniently, for identical orbitals the kinetic energy $T$ (together with one of potentials) cancels in the expression for the splitting, so the procedure only involves calculation of $S$ and potential matrix elements $\matrixel{1}{U_1}{2}$,  and $\matrixel{1}{U_2}{1}$, with which $\alpha$ and $\beta$ may thus be replaced, respectively.
	For non-degenerate orbitals, we also replace $\alpha'$ with $\matrixel{2}{U_1}{2}$.
	While in this case these replacements are not exact, they correspond to dropping the term with orbital energy difference in the result for the splitting of eigenstates, i.e., assures we calculate $2t$ instead of the full splitting $\sqrt{\lr{\Delta E}^2+4t^2}$.
	The result is
	\begin{equation}
		t = \frac{ \sqrt{ \cramped{ \lr*{s\beta^{*}+s^{*}\beta-\alpha-\alpha'}^2 - 4\lr*{\alpha\alpha'-\abs{\beta}^2}\lr*{1-\abs{S}^2} } } }{2\lr*{1-\abs{S}^2}} ,
	\end{equation}
	and reduces to just $t =(\beta-\alpha S)/(1-S^2)$ for real integrals and $\alpha=\alpha'$.
	
	We evaluate the above integrals using the same tools and methods as for the approach based on Bardeen's theory.
}

\section{Calculation details}\label{sec:details}
	Wave functions are constructed based on data read from supplemental files provided with \oncite{GamblePRB2015}.
	The levels belonging to the $T_2$ and $E$ manifolds are degenerate, so one deals with three- and two-dimensional eigenspaces, respectively. To obtain the specific $T_2^{(x)}$, $T_2^{(y)}$, $T_2^{(z)}$, $E^{(xy)}$, and $E^{(z)}$ states, we use linear combinations of the wave functions provided in \oncite{GamblePRB2015} that have the appropriate symmetry properties,
	\begin{align*}
		\ket[\big]{\,T_2^{(x)}} =& \phantom{-.} 0.674807 \,\,\ket[\big]{\,T_2^{(1)}} + 0.161526 \,\,\ket[\big]{\,T_2^{(2)}} \\
		&+ 0.720100 \,\,\ket[\big]{\,T_2^{(3)}},\nonumber\\
		\ket[\big]{\,T_2^{(y)}} =& -0.351186 \,\,\ket[\big]{\,T_2^{(1)}} + 0.928476 \,\,\ket[\big]{\,T_2^{(2)}}\\\
		&+ 0.120830 \,\,\ket[\big]{\,T_2^{(3)}},\nonumber\\
		\ket[\big]{\,T_2^{(z)}} =& -0.649079 \,\,\ket[\big]{\,T_2^{(1)}} - 0.334427 \,\,\ket[\big]{\,T_2^{(2)}}\\
		&+ 0.683268 \,\,\ket[\big]{\,T_2^{(3)}},\nonumber\\
		\ket[\big]{\,E^{(xy)}}  =& \phantom{-.} 0.707120 \,\,\ket[\big]{\,E^{(1)}}   + 0.707094 \,\,\ket[\big]{\,E^{(2)}},\nonumber\\
		\ket[\big]{\,E^{(z)}}   =& -0.707094 \,\,\ket[\big]{\,E^{(1)}}   + 0.707120  \,\,\ket[\big]{\,E^{(2)}}.\nonumber
	\end{align*}
	Numerical integration is done using the Vegas algorithm \cite{PeterLepageJCP1978} of adaptive Monte Carlo integration implemented in the Cuba library \cite{HahnCPC2005}.

\section{Computational cost}\label{sec:calctime}
	\begin{figure}[tb!]
		\includegraphics[width=1.0\linewidth]{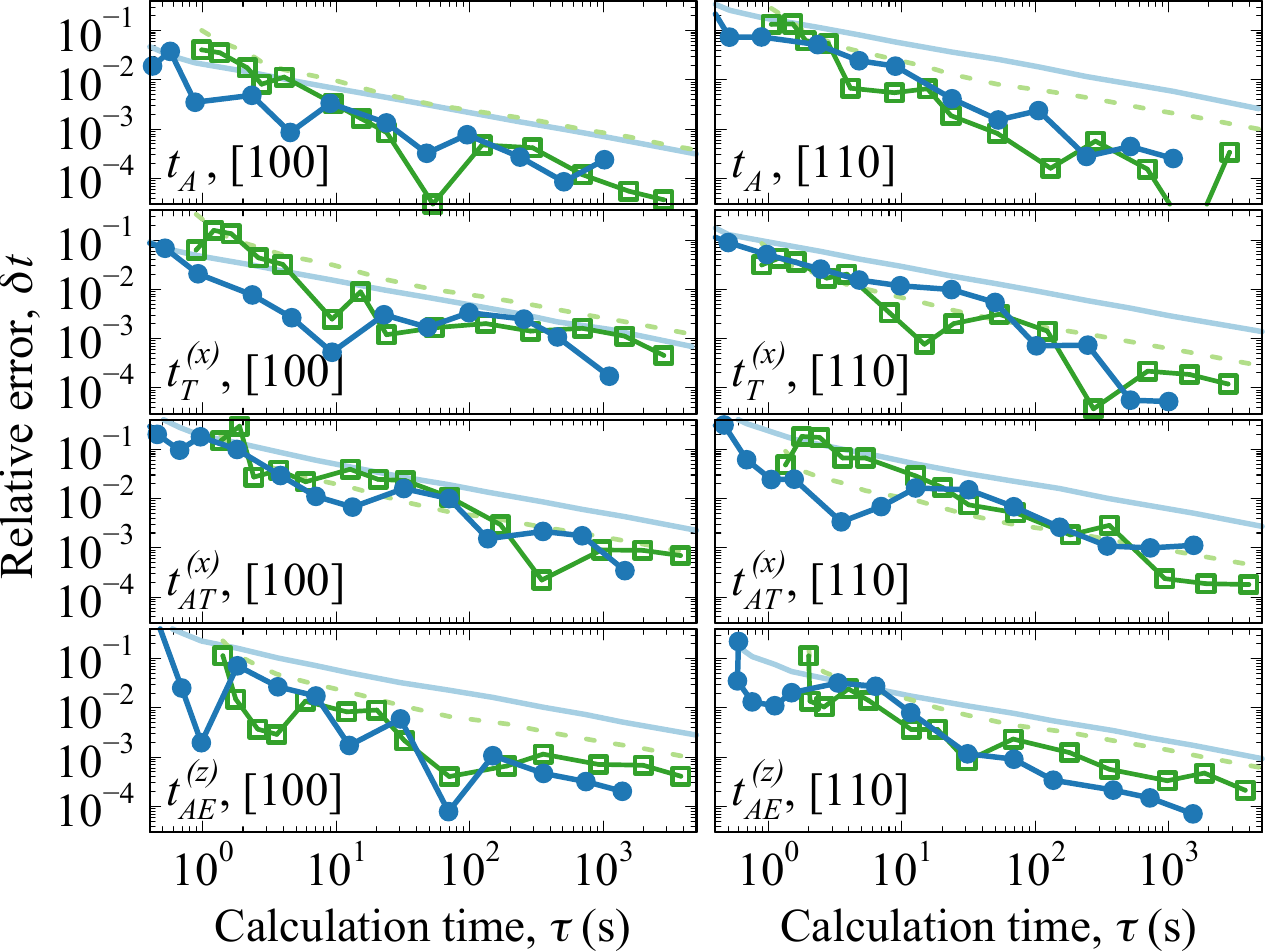} %
		\caption{\label{fig:calctime}(Color online) Estimated error versus calculation time. The relative calculation error as reported by the integration algorithm (lines) and the actual error relative to the converged values (symbols) plotted as a function of calculation time. Full circles and solid lines show values for the presented method, while empty squares and dashed lines are for the standard one. Each panel presents results for a different orbital pair and displacement axis, as marked. Lines connecting points are to guide the eye.} %
	\end{figure}

	Here, we evaluate the computational cost of the proposed method.
	For this, we take fixed and possibly similar distances along [100] and [110], $d=4.34456$~nm and $d=4.22409$~nm, respectively.
	First, we converge the results up to the maximal feasible accuracy.
	For this, we use a fixed number of integrand evaluations of $10^9$.
	Next, we notice that the error estimation by the integration procedure is typically significantly larger than the actual difference of the result compared to the converged one.
	Of course, the converged result is not a strict reference, as it is not exact, but its possible error is orders of magnitude smaller than the differences in question, which justifies such a treatment.
	The error overestimation by the integration procedure is understandable, as it reports the upper bound that can be only approximately estimated.
	For a figure of merit, we take the actual difference of a result relative to the converged one.
	To evaluate the dependence on computation time, we 
	perform a series of calculations with an increasing number of integrand evaluations.
	\figref{calctime} shows the relative errors calculated this way with respect to the converged values, with full and open symbols for the presented and standard methods, respectively.
	With solid and dashed lines, we also show the upper bound for the error as reported by the integration procedure.
	While it is typically lower for the standard method, one may notice that the actual convergence of both methods is similar.
	Thus, considering the conceptual advantages of Bardeen's theory, we may conclude that the proposed method is at least competitive to the standard treatment.


\begin{thebibliography}{53}%
	\makeatletter
	\providecommand \@ifxundefined [1]{%
		\@ifx{#1\undefined}
	}%
	\providecommand \@ifnum [1]{%
		\ifnum #1\expandafter \@firstoftwo
		\else \expandafter \@secondoftwo
		\fi
	}%
	\providecommand \@ifx [1]{%
		\ifx #1\expandafter \@firstoftwo
		\else \expandafter \@secondoftwo
		\fi
	}%
	\providecommand \natexlab [1]{#1}%
	\providecommand \enquote  [1]{``#1''}%
	\providecommand \bibnamefont  [1]{#1}%
	\providecommand \bibfnamefont [1]{#1}%
	\providecommand \citenamefont [1]{#1}%
	\providecommand \href@noop [0]{\@secondoftwo}%
	\providecommand \href [0]{\begingroup \@sanitize@url \@href}%
	\providecommand \@href[1]{\@@startlink{#1}\@@href}%
	\providecommand \@@href[1]{\endgroup#1\@@endlink}%
	\providecommand \@sanitize@url [0]{\catcode `\\12\catcode `\$12\catcode
		`\&12\catcode `\#12\catcode `\^12\catcode `\_12\catcode `\%12\relax}%
	\providecommand \@@startlink[1]{}%
	\providecommand \@@endlink[0]{}%
	\providecommand \url  [0]{\begingroup\@sanitize@url \@url }%
	\providecommand \@url [1]{\endgroup\@href {#1}{\urlprefix }}%
	\providecommand \urlprefix  [0]{URL }%
	\providecommand \Eprint [0]{\href }%
	\providecommand \doibase [0]{https://doi.org/}%
	\providecommand \selectlanguage [0]{\@gobble}%
	\providecommand \bibinfo  [0]{\@secondoftwo}%
	\providecommand \bibfield  [0]{\@secondoftwo}%
	\providecommand \translation [1]{[#1]}%
	\providecommand \BibitemOpen [0]{}%
	\providecommand \bibitemStop [0]{}%
	\providecommand \bibitemNoStop [0]{.\EOS\space}%
	\providecommand \EOS [0]{\spacefactor3000\relax}%
	\providecommand \BibitemShut  [1]{\csname bibitem#1\endcsname}%
	\let\auto@bib@innerbib\@empty
	\bibitem [{\citenamefont {Kane}(1998)}]{KaneN1998}%
	\BibitemOpen
	\bibfield  {author} {\bibinfo {author} {\bibfnamefont {B.~E.}\ \bibnamefont
			{Kane}},\ }\href {https://doi.org/10.1038/30156} {\bibfield  {journal}
		{\bibinfo  {journal} {Nature}\ }\textbf {\bibinfo {volume} {393}},\ \bibinfo
		{pages} {133} (\bibinfo {year} {1998})}\BibitemShut {NoStop}%
	\bibitem [{\citenamefont {Vrijen}\ \emph {et~al.}(2000)\citenamefont {Vrijen},
		\citenamefont {Yablonovitch}, \citenamefont {Wang}, \citenamefont {Jiang},
		\citenamefont {Balandin}, \citenamefont {Roychowdhury}, \citenamefont {Mor},\
		and\ \citenamefont {DiVincenzo}}]{VrijenPRA2000}%
	\BibitemOpen
	\bibfield  {author} {\bibinfo {author} {\bibfnamefont {R.}~\bibnamefont
			{Vrijen}}, \bibinfo {author} {\bibfnamefont {E.}~\bibnamefont
			{Yablonovitch}}, \bibinfo {author} {\bibfnamefont {K.}~\bibnamefont {Wang}},
		\bibinfo {author} {\bibfnamefont {H.~W.}\ \bibnamefont {Jiang}}, \bibinfo
		{author} {\bibfnamefont {A.}~\bibnamefont {Balandin}}, \bibinfo {author}
		{\bibfnamefont {V.}~\bibnamefont {Roychowdhury}}, \bibinfo {author}
		{\bibfnamefont {T.}~\bibnamefont {Mor}},\ and\ \bibinfo {author}
		{\bibfnamefont {D.}~\bibnamefont {DiVincenzo}},\ }\href
	{https://doi.org/10.1103/PhysRevA.62.012306} {\bibfield  {journal} {\bibinfo
			{journal} {Phys. Rev. A}\ }\textbf {\bibinfo {volume} {62}},\ \bibinfo
		{pages} {012306} (\bibinfo {year} {2000})}\BibitemShut {NoStop}%
	\bibitem [{\citenamefont {Hollenberg}\ \emph {et~al.}(2004)\citenamefont
		{Hollenberg}, \citenamefont {Dzurak}, \citenamefont {Wellard}, \citenamefont
		{Hamilton}, \citenamefont {Reilly}, \citenamefont {Milburn},\ and\
		\citenamefont {Clark}}]{HollenbergPRB2004}%
	\BibitemOpen
	\bibfield  {author} {\bibinfo {author} {\bibfnamefont {L.~C.~L.}\
			\bibnamefont {Hollenberg}}, \bibinfo {author} {\bibfnamefont {A.~S.}\
			\bibnamefont {Dzurak}}, \bibinfo {author} {\bibfnamefont {C.}~\bibnamefont
			{Wellard}}, \bibinfo {author} {\bibfnamefont {A.~R.}\ \bibnamefont
			{Hamilton}}, \bibinfo {author} {\bibfnamefont {D.~J.}\ \bibnamefont
			{Reilly}}, \bibinfo {author} {\bibfnamefont {G.~J.}\ \bibnamefont
			{Milburn}},\ and\ \bibinfo {author} {\bibfnamefont {R.~G.}\ \bibnamefont
			{Clark}},\ }\href {https://doi.org/10.1103/PhysRevB.69.113301} {\bibfield
		{journal} {\bibinfo  {journal} {Phys. Rev. B}\ }\textbf {\bibinfo {volume}
			{69}},\ \bibinfo {pages} {113301} (\bibinfo {year} {2004})}\BibitemShut
	{NoStop}%
	\bibitem [{\citenamefont {O'Brien}\ \emph {et~al.}(2001)\citenamefont
		{O'Brien}, \citenamefont {Schofield}, \citenamefont {Simmons}, \citenamefont
		{Clark}, \citenamefont {Dzurak}, \citenamefont {Curson}, \citenamefont
		{Kane}, \citenamefont {McAlpine}, \citenamefont {Hawley},\ and\ \citenamefont
		{Brown}}]{OBrienPRB2001}%
	\BibitemOpen
	\bibfield  {author} {\bibinfo {author} {\bibfnamefont {J.~L.}\ \bibnamefont
			{O'Brien}}, \bibinfo {author} {\bibfnamefont {S.~R.}\ \bibnamefont
			{Schofield}}, \bibinfo {author} {\bibfnamefont {M.~Y.}\ \bibnamefont
			{Simmons}}, \bibinfo {author} {\bibfnamefont {R.~G.}\ \bibnamefont {Clark}},
		\bibinfo {author} {\bibfnamefont {A.~S.}\ \bibnamefont {Dzurak}}, \bibinfo
		{author} {\bibfnamefont {N.~J.}\ \bibnamefont {Curson}}, \bibinfo {author}
		{\bibfnamefont {B.~E.}\ \bibnamefont {Kane}}, \bibinfo {author}
		{\bibfnamefont {N.~S.}\ \bibnamefont {McAlpine}}, \bibinfo {author}
		{\bibfnamefont {M.~E.}\ \bibnamefont {Hawley}},\ and\ \bibinfo {author}
		{\bibfnamefont {G.~W.}\ \bibnamefont {Brown}},\ }\href
	{https://doi.org/10.1103/PhysRevB.64.161401} {\bibfield  {journal} {\bibinfo
			{journal} {Phys. Rev. B}\ }\textbf {\bibinfo {volume} {64}},\ \bibinfo
		{pages} {161401(R)} (\bibinfo {year} {2001})}\BibitemShut {NoStop}%
	\bibitem [{\citenamefont {Pla}\ \emph {et~al.}(2013)\citenamefont {Pla},
		\citenamefont {Tan}, \citenamefont {Dehollain}, \citenamefont {Lim},
		\citenamefont {Morton}, \citenamefont {Zwanenburg}, \citenamefont {Jamieson},
		\citenamefont {Dzurak},\ and\ \citenamefont {Morello}}]{JarrydN2013}%
	\BibitemOpen
	\bibfield  {author} {\bibinfo {author} {\bibfnamefont {J.~J.}\ \bibnamefont
			{Pla}}, \bibinfo {author} {\bibfnamefont {K.~Y.}\ \bibnamefont {Tan}},
		\bibinfo {author} {\bibfnamefont {J.~P.}\ \bibnamefont {Dehollain}}, \bibinfo
		{author} {\bibfnamefont {W.~H.}\ \bibnamefont {Lim}}, \bibinfo {author}
		{\bibfnamefont {J.~J.~L.}\ \bibnamefont {Morton}}, \bibinfo {author}
		{\bibfnamefont {F.~A.}\ \bibnamefont {Zwanenburg}}, \bibinfo {author}
		{\bibfnamefont {D.~N.}\ \bibnamefont {Jamieson}}, \bibinfo {author}
		{\bibfnamefont {A.~S.}\ \bibnamefont {Dzurak}},\ and\ \bibinfo {author}
		{\bibfnamefont {A.}~\bibnamefont {Morello}},\ }\href
	{https://doi.org/10.1038/nature12011} {\bibfield  {journal} {\bibinfo
			{journal} {Nature}\ }\textbf {\bibinfo {volume} {496}},\ \bibinfo {pages}
		{334} (\bibinfo {year} {2013})}\BibitemShut {NoStop}%
	\bibitem [{\citenamefont {Dehollain}\ \emph {et~al.}(2014)\citenamefont
		{Dehollain}, \citenamefont {Muhonen}, \citenamefont {Tan}, \citenamefont
		{Saraiva}, \citenamefont {Jamieson}, \citenamefont {Dzurak},\ and\
		\citenamefont {Morello}}]{DehollainPRL2014}%
	\BibitemOpen
	\bibfield  {author} {\bibinfo {author} {\bibfnamefont {J.~P.}\ \bibnamefont
			{Dehollain}}, \bibinfo {author} {\bibfnamefont {J.~T.}\ \bibnamefont
			{Muhonen}}, \bibinfo {author} {\bibfnamefont {K.~Y.}\ \bibnamefont {Tan}},
		\bibinfo {author} {\bibfnamefont {A.}~\bibnamefont {Saraiva}}, \bibinfo
		{author} {\bibfnamefont {D.~N.}\ \bibnamefont {Jamieson}}, \bibinfo {author}
		{\bibfnamefont {A.~S.}\ \bibnamefont {Dzurak}},\ and\ \bibinfo {author}
		{\bibfnamefont {A.}~\bibnamefont {Morello}},\ }\href
	{https://doi.org/10.1103/PhysRevLett.112.236801} {\bibfield  {journal}
		{\bibinfo  {journal} {Phys. Rev. Lett.}\ }\textbf {\bibinfo {volume} {112}},\
		\bibinfo {pages} {236801} (\bibinfo {year} {2014})}\BibitemShut {NoStop}%
	\bibitem [{\citenamefont {Gonzalez-Zalba}\ \emph {et~al.}(2014)\citenamefont
		{Gonzalez-Zalba}, \citenamefont {Saraiva}, \citenamefont {Calder{\'{o}}n},
		\citenamefont {Heiss}, \citenamefont {Koiller},\ and\ \citenamefont
		{Ferguson}}]{GonzalezNL2014}%
	\BibitemOpen
	\bibfield  {author} {\bibinfo {author} {\bibfnamefont {M.~F.}\ \bibnamefont
			{Gonzalez-Zalba}}, \bibinfo {author} {\bibfnamefont {A.}~\bibnamefont
			{Saraiva}}, \bibinfo {author} {\bibfnamefont {M.~J.}\ \bibnamefont
			{Calder{\'{o}}n}}, \bibinfo {author} {\bibfnamefont {D.}~\bibnamefont
			{Heiss}}, \bibinfo {author} {\bibfnamefont {B.}~\bibnamefont {Koiller}},\
		and\ \bibinfo {author} {\bibfnamefont {A.~J.}\ \bibnamefont {Ferguson}},\
	}\href {https://doi.org/10.1021/nl5023942} {\bibfield  {journal} {\bibinfo
			{journal} {Nano Lett.}\ }\textbf {\bibinfo {volume} {14}},\ \bibinfo {pages}
		{5672} (\bibinfo {year} {2014})}\BibitemShut {NoStop}%
	\bibitem [{\citenamefont {Fuechsle}\ \emph {et~al.}(2012)\citenamefont
		{Fuechsle}, \citenamefont {Miwa}, \citenamefont {Mahapatra}, \citenamefont
		{Ryu}, \citenamefont {Lee}, \citenamefont {Warschkow}, \citenamefont
		{Hollenberg}, \citenamefont {Klimeck},\ and\ \citenamefont
		{Simmons}}]{FuechsleNN2012}%
	\BibitemOpen
	\bibfield  {author} {\bibinfo {author} {\bibfnamefont {M.}~\bibnamefont
			{Fuechsle}}, \bibinfo {author} {\bibfnamefont {J.~A.}\ \bibnamefont {Miwa}},
		\bibinfo {author} {\bibfnamefont {S.}~\bibnamefont {Mahapatra}}, \bibinfo
		{author} {\bibfnamefont {H.}~\bibnamefont {Ryu}}, \bibinfo {author}
		{\bibfnamefont {S.}~\bibnamefont {Lee}}, \bibinfo {author} {\bibfnamefont
			{O.}~\bibnamefont {Warschkow}}, \bibinfo {author} {\bibfnamefont {L.~C.~L.}\
			\bibnamefont {Hollenberg}}, \bibinfo {author} {\bibfnamefont
			{G.}~\bibnamefont {Klimeck}},\ and\ \bibinfo {author} {\bibfnamefont {M.~Y.}\
			\bibnamefont {Simmons}},\ }\href {https://doi.org/10.1038/nnano.2012.21}
	{\bibfield  {journal} {\bibinfo  {journal} {Nature Nanotech.}\ }\textbf
		{\bibinfo {volume} {7}},\ \bibinfo {pages} {242} (\bibinfo {year}
		{2012})}\BibitemShut {NoStop}%
	\bibitem [{\citenamefont {B\"{u}ch}\ \emph {et~al.}(2013)\citenamefont
		{B\"{u}ch}, \citenamefont {Mahapatra}, \citenamefont {Rahman}, \citenamefont
		{Morello},\ and\ \citenamefont {Simmons}}]{BuchNC2013}%
	\BibitemOpen
	\bibfield  {author} {\bibinfo {author} {\bibfnamefont {H.}~\bibnamefont
			{B\"{u}ch}}, \bibinfo {author} {\bibfnamefont {S.}~\bibnamefont {Mahapatra}},
		\bibinfo {author} {\bibfnamefont {R.}~\bibnamefont {Rahman}}, \bibinfo
		{author} {\bibfnamefont {A.}~\bibnamefont {Morello}},\ and\ \bibinfo {author}
		{\bibfnamefont {M.~Y.}\ \bibnamefont {Simmons}},\ }\href
	{https://doi.org/10.1038/ncomms3017} {\bibfield  {journal} {\bibinfo
			{journal} {Nature Commun.}\ }\textbf {\bibinfo {volume} {4}},\ \bibinfo
		{pages} {2017} (\bibinfo {year} {2013})}\BibitemShut {NoStop}%
	\bibitem [{\citenamefont {Wyrick}\ \emph {et~al.}(2019)\citenamefont {Wyrick},
		\citenamefont {Wang}, \citenamefont {Kashid}, \citenamefont {Namboodiri},
		\citenamefont {Schmucker}, \citenamefont {Hagmann}, \citenamefont {Liu},
		\citenamefont {Stewart}, \citenamefont {Richter}, \citenamefont {Bryant},\
		and\ \citenamefont {Silver}}]{WyrickAFM2019}%
	\BibitemOpen
	\bibfield  {author} {\bibinfo {author} {\bibfnamefont {J.}~\bibnamefont
			{Wyrick}}, \bibinfo {author} {\bibfnamefont {X.}~\bibnamefont {Wang}},
		\bibinfo {author} {\bibfnamefont {R.~V.}\ \bibnamefont {Kashid}}, \bibinfo
		{author} {\bibfnamefont {P.}~\bibnamefont {Namboodiri}}, \bibinfo {author}
		{\bibfnamefont {S.~W.}\ \bibnamefont {Schmucker}}, \bibinfo {author}
		{\bibfnamefont {J.~A.}\ \bibnamefont {Hagmann}}, \bibinfo {author}
		{\bibfnamefont {K.}~\bibnamefont {Liu}}, \bibinfo {author} {\bibfnamefont
			{M.~D.}\ \bibnamefont {Stewart}}, \bibinfo {author} {\bibfnamefont {C.~A.}\
			\bibnamefont {Richter}}, \bibinfo {author} {\bibfnamefont {G.~W.}\
			\bibnamefont {Bryant}},\ and\ \bibinfo {author} {\bibfnamefont {R.~M.}\
			\bibnamefont {Silver}},\ }\href {https://doi.org/10.1002/adfm.201903475}
	{\bibfield  {journal} {\bibinfo  {journal} {Adv. Funct. Mater.}\ }\textbf
		{\bibinfo {volume} {29}},\ \bibinfo {pages} {1903475} (\bibinfo {year}
		{2019})}\BibitemShut {NoStop}%
	\bibitem [{\citenamefont {Wang}\ \emph {et~al.}(2020)\citenamefont {Wang},
		\citenamefont {Wyrick}, \citenamefont {Kashid}, \citenamefont {Namboodiri},
		\citenamefont {Schmucker}, \citenamefont {Murphy}, \citenamefont {Stewart},\
		and\ \citenamefont {Silver}}]{WangCP2020}%
	\BibitemOpen
	\bibfield  {author} {\bibinfo {author} {\bibfnamefont {X.}~\bibnamefont
			{Wang}}, \bibinfo {author} {\bibfnamefont {J.}~\bibnamefont {Wyrick}},
		\bibinfo {author} {\bibfnamefont {R.~V.}\ \bibnamefont {Kashid}}, \bibinfo
		{author} {\bibfnamefont {P.}~\bibnamefont {Namboodiri}}, \bibinfo {author}
		{\bibfnamefont {S.~W.}\ \bibnamefont {Schmucker}}, \bibinfo {author}
		{\bibfnamefont {A.}~\bibnamefont {Murphy}}, \bibinfo {author} {\bibfnamefont
			{M.~D.}\ \bibnamefont {Stewart}},\ and\ \bibinfo {author} {\bibfnamefont
			{R.~M.}\ \bibnamefont {Silver}},\ }\href
	{https://doi.org/10.1038/s42005-020-0343-1} {\bibfield  {journal} {\bibinfo
			{journal} {Commun. Phys.}\ }\textbf {\bibinfo {volume} {3}},\ \bibinfo
		{pages} {82} (\bibinfo {year} {2020})}\BibitemShut {NoStop}%
	\bibitem [{\citenamefont {Zwanenburg}\ \emph {et~al.}(2013)\citenamefont
		{Zwanenburg}, \citenamefont {Dzurak}, \citenamefont {Morello}, \citenamefont
		{Simmons}, \citenamefont {Hollenberg}, \citenamefont {Klimeck}, \citenamefont
		{Rogge}, \citenamefont {Coppersmith},\ and\ \citenamefont
		{Eriksson}}]{ZwanenburgRoMP2013}%
	\BibitemOpen
	\bibfield  {author} {\bibinfo {author} {\bibfnamefont {F.~A.}\ \bibnamefont
			{Zwanenburg}}, \bibinfo {author} {\bibfnamefont {A.~S.}\ \bibnamefont
			{Dzurak}}, \bibinfo {author} {\bibfnamefont {A.}~\bibnamefont {Morello}},
		\bibinfo {author} {\bibfnamefont {M.~Y.}\ \bibnamefont {Simmons}}, \bibinfo
		{author} {\bibfnamefont {L.~C.~L.}\ \bibnamefont {Hollenberg}}, \bibinfo
		{author} {\bibfnamefont {G.}~\bibnamefont {Klimeck}}, \bibinfo {author}
		{\bibfnamefont {S.}~\bibnamefont {Rogge}}, \bibinfo {author} {\bibfnamefont
			{S.~N.}\ \bibnamefont {Coppersmith}},\ and\ \bibinfo {author} {\bibfnamefont
			{M.~A.}\ \bibnamefont {Eriksson}},\ }\href
	{https://doi.org/10.1103/revmodphys.85.961} {\bibfield  {journal} {\bibinfo
			{journal} {Rev. Mod. Phys.}\ }\textbf {\bibinfo {volume} {85}},\ \bibinfo
		{pages} {961} (\bibinfo {year} {2013})}\BibitemShut {NoStop}%
	\bibitem [{\citenamefont {Alipour}\ \emph {et~al.}(2022)\citenamefont
		{Alipour}, \citenamefont {Fowler}, \citenamefont {Moheimani}, \citenamefont
		{Owen},\ and\ \citenamefont {Randall}}]{AlipourJVS2022}%
	\BibitemOpen
	\bibfield  {author} {\bibinfo {author} {\bibfnamefont {A.}~\bibnamefont
			{Alipour}}, \bibinfo {author} {\bibfnamefont {E.~L.}\ \bibnamefont {Fowler}},
		\bibinfo {author} {\bibfnamefont {S.~O.~R.}\ \bibnamefont {Moheimani}},
		\bibinfo {author} {\bibfnamefont {J.~H.~G.}\ \bibnamefont {Owen}},\ and\
		\bibinfo {author} {\bibfnamefont {J.~N.}\ \bibnamefont {Randall}},\ }\href
	{https://doi.org/10.1116/6.0001826} {\bibfield  {journal} {\bibinfo
			{journal} {Journal of Vacuum Science \& Technology B}\ }\textbf {\bibinfo
			{volume} {40}},\ \bibinfo {pages} {030603} (\bibinfo {year}
		{2022})}\BibitemShut {NoStop}%
	\bibitem [{\citenamefont {Wang}\ \emph {et~al.}(2021)\citenamefont {Wang},
		\citenamefont {Khatami}, \citenamefont {Fei}, \citenamefont {Wyrick},
		\citenamefont {Namboodiri}, \citenamefont {Kashid}, \citenamefont {Rigosi},
		\citenamefont {Bryant},\ and\ \citenamefont {Silver}}]{Wang2021}%
	\BibitemOpen
	\bibfield  {author} {\bibinfo {author} {\bibfnamefont {X.}~\bibnamefont
			{Wang}}, \bibinfo {author} {\bibfnamefont {E.}~\bibnamefont {Khatami}},
		\bibinfo {author} {\bibfnamefont {F.}~\bibnamefont {Fei}}, \bibinfo {author}
		{\bibfnamefont {J.}~\bibnamefont {Wyrick}}, \bibinfo {author} {\bibfnamefont
			{P.}~\bibnamefont {Namboodiri}}, \bibinfo {author} {\bibfnamefont
			{R.}~\bibnamefont {Kashid}}, \bibinfo {author} {\bibfnamefont {A.~F.}\
			\bibnamefont {Rigosi}}, \bibinfo {author} {\bibfnamefont {G.}~\bibnamefont
			{Bryant}},\ and\ \bibinfo {author} {\bibfnamefont {R.}~\bibnamefont
			{Silver}},\ }\href@noop {} {\bibinfo {title} {Quantum simulation of an
			extended fermi-hubbard model using a {2D} lattice of dopant-based quantum
			dots}} (\bibinfo {year} {2021}),\ \Eprint
	{https://arxiv.org/abs/arXiv:2110.08982} {arXiv:2110.08982} \BibitemShut
	{NoStop}%
	\bibitem [{\citenamefont {Kohn}\ and\ \citenamefont
		{Luttinger}(1955)}]{KohnPR1955}%
	\BibitemOpen
	\bibfield  {author} {\bibinfo {author} {\bibfnamefont {W.}~\bibnamefont
			{Kohn}}\ and\ \bibinfo {author} {\bibfnamefont {J.~M.}\ \bibnamefont
			{Luttinger}},\ }\href {https://doi.org/10.1103/PhysRev.98.915} {\bibfield
		{journal} {\bibinfo  {journal} {Phys. Rev.}\ }\textbf {\bibinfo {volume}
			{98}},\ \bibinfo {pages} {915} (\bibinfo {year} {1955})}\BibitemShut
	{NoStop}%
	\bibitem [{\citenamefont {Klymenko}\ and\ \citenamefont
		{Remacle}(2014)}]{KlymenkoJPCM2014}%
	\BibitemOpen
	\bibfield  {author} {\bibinfo {author} {\bibfnamefont {M.~V.}\ \bibnamefont
			{Klymenko}}\ and\ \bibinfo {author} {\bibfnamefont {F.}~\bibnamefont
			{Remacle}},\ }\href {https://doi.org/10.1088/0953-8984/26/6/065302}
	{\bibfield  {journal} {\bibinfo  {journal} {J. Phys.: Condens. Matter}\
		}\textbf {\bibinfo {volume} {26}},\ \bibinfo {pages} {065302} (\bibinfo
		{year} {2014})}\BibitemShut {NoStop}%
	\bibitem [{\citenamefont {Saraiva}\ \emph {et~al.}(2015)\citenamefont
		{Saraiva}, \citenamefont {Baena}, \citenamefont {Calder{\'{o}}n},\ and\
		\citenamefont {Koiller}}]{SaraivaJPCM2015}%
	\BibitemOpen
	\bibfield  {author} {\bibinfo {author} {\bibfnamefont {A.~L.}\ \bibnamefont
			{Saraiva}}, \bibinfo {author} {\bibfnamefont {A.}~\bibnamefont {Baena}},
		\bibinfo {author} {\bibfnamefont {M.~J.}\ \bibnamefont {Calder{\'{o}}n}},\
		and\ \bibinfo {author} {\bibfnamefont {B.}~\bibnamefont {Koiller}},\ }\href
	{https://doi.org/10.1088/0953-8984/27/15/154208} {\bibfield  {journal}
		{\bibinfo  {journal} {J. Phys.: Condens. Matter}\ }\textbf {\bibinfo {volume}
			{27}},\ \bibinfo {pages} {154208} (\bibinfo {year} {2015})}\BibitemShut
	{NoStop}%
	\bibitem [{\citenamefont {Gamble}\ \emph {et~al.}(2015)\citenamefont {Gamble},
		\citenamefont {Jacobson}, \citenamefont {Nielsen}, \citenamefont {Baczewski},
		\citenamefont {Moussa}, \citenamefont {Monta{\~{n}}o},\ and\ \citenamefont
		{Muller}}]{GamblePRB2015}%
	\BibitemOpen
	\bibfield  {author} {\bibinfo {author} {\bibfnamefont {J.~K.}\ \bibnamefont
			{Gamble}}, \bibinfo {author} {\bibfnamefont {N.~T.}\ \bibnamefont
			{Jacobson}}, \bibinfo {author} {\bibfnamefont {E.}~\bibnamefont {Nielsen}},
		\bibinfo {author} {\bibfnamefont {A.~D.}\ \bibnamefont {Baczewski}}, \bibinfo
		{author} {\bibfnamefont {J.~E.}\ \bibnamefont {Moussa}}, \bibinfo {author}
		{\bibfnamefont {I.}~\bibnamefont {Monta\~no}},\ and\ \bibinfo {author}
		{\bibfnamefont {R.~P.}\ \bibnamefont {Muller}},\ }\href
	{https://doi.org/10.1103/physrevb.91.235318} {\bibfield  {journal} {\bibinfo
			{journal} {Phys. Rev. B}\ }\textbf {\bibinfo {volume} {91}},\ \bibinfo
		{pages} {235318} (\bibinfo {year} {2015})}\BibitemShut {NoStop}%
	\bibitem [{\citenamefont {Menchero}\ \emph {et~al.}(1999)\citenamefont
		{Menchero}, \citenamefont {Capaz}, \citenamefont {Koiller},\ and\
		\citenamefont {Chacham}}]{MencheroPRB1999}%
	\BibitemOpen
	\bibfield  {author} {\bibinfo {author} {\bibfnamefont {J.~G.}\ \bibnamefont
			{Menchero}}, \bibinfo {author} {\bibfnamefont {R.~B.}\ \bibnamefont {Capaz}},
		\bibinfo {author} {\bibfnamefont {B.}~\bibnamefont {Koiller}},\ and\ \bibinfo
		{author} {\bibfnamefont {H.}~\bibnamefont {Chacham}},\ }\href
	{https://doi.org/10.1103/PhysRevB.59.2722} {\bibfield  {journal} {\bibinfo
			{journal} {Phys. Rev. B}\ }\textbf {\bibinfo {volume} {59}},\ \bibinfo
		{pages} {2722} (\bibinfo {year} {1999})}\BibitemShut {NoStop}%
	\bibitem [{\citenamefont {Martins}\ \emph {et~al.}(2005)\citenamefont
		{Martins}, \citenamefont {Boykin}, \citenamefont {Klimeck},\ and\
		\citenamefont {Koiller}}]{MartinsPRB2005}%
	\BibitemOpen
	\bibfield  {author} {\bibinfo {author} {\bibfnamefont {A.~S.}\ \bibnamefont
			{Martins}}, \bibinfo {author} {\bibfnamefont {T.~B.}\ \bibnamefont {Boykin}},
		\bibinfo {author} {\bibfnamefont {G.}~\bibnamefont {Klimeck}},\ and\ \bibinfo
		{author} {\bibfnamefont {B.}~\bibnamefont {Koiller}},\ }\href
	{https://doi.org/10.1103/PhysRevB.72.193204} {\bibfield  {journal} {\bibinfo
			{journal} {Phys. Rev. B}\ }\textbf {\bibinfo {volume} {72}},\ \bibinfo
		{pages} {193204} (\bibinfo {year} {2005})}\BibitemShut {NoStop}%
	\bibitem [{\citenamefont {Klimeck}\ \emph {et~al.}(2002)\citenamefont
		{Klimeck}, \citenamefont {Oyafuso}, \citenamefont {Boykin}, \citenamefont
		{Bowen},\ and\ \citenamefont {von Allmen}}]{KlimeckCMiES2002}%
	\BibitemOpen
	\bibfield  {author} {\bibinfo {author} {\bibfnamefont {G.}~\bibnamefont
			{Klimeck}}, \bibinfo {author} {\bibfnamefont {F.}~\bibnamefont {Oyafuso}},
		\bibinfo {author} {\bibfnamefont {T.~B.}\ \bibnamefont {Boykin}}, \bibinfo
		{author} {\bibfnamefont {R.~C.}\ \bibnamefont {Bowen}},\ and\ \bibinfo
		{author} {\bibfnamefont {P.}~\bibnamefont {von Allmen}},\ }\href
	{https://doi.org/10.3970/cmes.2002.003.601} {\bibfield  {journal} {\bibinfo
			{journal} {Comput. Model. Eng. Sci.}\ }\textbf {\bibinfo {volume} {3}},\
		\bibinfo {pages} {601} (\bibinfo {year} {2002})}\BibitemShut {NoStop}%
	\bibitem [{\citenamefont {Rahman}\ \emph {et~al.}(2011)\citenamefont {Rahman},
		\citenamefont {Park}, \citenamefont {Klimeck},\ and\ \citenamefont
		{Hollenberg}}]{RahmanNano2011}%
	\BibitemOpen
	\bibfield  {author} {\bibinfo {author} {\bibfnamefont {R.}~\bibnamefont
			{Rahman}}, \bibinfo {author} {\bibfnamefont {S.~H.}\ \bibnamefont {Park}},
		\bibinfo {author} {\bibfnamefont {G.}~\bibnamefont {Klimeck}},\ and\ \bibinfo
		{author} {\bibfnamefont {L.~C.~L.}\ \bibnamefont {Hollenberg}},\ }\href
	{https://doi.org/10.1088/0957-4484/22/22/225202} {\bibfield  {journal}
		{\bibinfo  {journal} {Nanotechnology}\ }\textbf {\bibinfo {volume} {22}},\
		\bibinfo {pages} {225202} (\bibinfo {year} {2011})}\BibitemShut {NoStop}%
	\bibitem [{\citenamefont {Tankasala}\ \emph {et~al.}(2022)\citenamefont
		{Tankasala}, \citenamefont {Voisin}, \citenamefont {Kembrey}, \citenamefont
		{Salfi}, \citenamefont {Hsueh}, \citenamefont {Osika}, \citenamefont
		{Rogge},\ and\ \citenamefont {Rahman}}]{TankasalaPRB2022}%
	\BibitemOpen
	\bibfield  {author} {\bibinfo {author} {\bibfnamefont {A.}~\bibnamefont
			{Tankasala}}, \bibinfo {author} {\bibfnamefont {B.}~\bibnamefont {Voisin}},
		\bibinfo {author} {\bibfnamefont {Z.}~\bibnamefont {Kembrey}}, \bibinfo
		{author} {\bibfnamefont {J.}~\bibnamefont {Salfi}}, \bibinfo {author}
		{\bibfnamefont {Y.-L.}\ \bibnamefont {Hsueh}}, \bibinfo {author}
		{\bibfnamefont {E.~N.}\ \bibnamefont {Osika}}, \bibinfo {author}
		{\bibfnamefont {S.}~\bibnamefont {Rogge}},\ and\ \bibinfo {author}
		{\bibfnamefont {R.}~\bibnamefont {Rahman}},\ }\href
	{https://doi.org/10.1103/PhysRevB.105.155158} {\bibfield  {journal} {\bibinfo
			{journal} {Phys. Rev. B}\ }\textbf {\bibinfo {volume} {105}},\ \bibinfo
		{pages} {155158} (\bibinfo {year} {2022})}\BibitemShut {NoStop}%
	\bibitem [{\citenamefont {Wu}\ and\ \citenamefont {Fisher}(2021)}]{WuPRB2021}%
	\BibitemOpen
	\bibfield  {author} {\bibinfo {author} {\bibfnamefont {W.}~\bibnamefont
			{Wu}}\ and\ \bibinfo {author} {\bibfnamefont {A.~J.}\ \bibnamefont
			{Fisher}},\ }\href {https://doi.org/10.1103/PhysRevB.104.035433} {\bibfield
		{journal} {\bibinfo  {journal} {Phys. Rev. B}\ }\textbf {\bibinfo {volume}
			{104}},\ \bibinfo {pages} {035433} (\bibinfo {year} {2021})}\BibitemShut
	{NoStop}%
	\bibitem [{\citenamefont {Jagannath}\ \emph {et~al.}(1981)\citenamefont
		{Jagannath}, \citenamefont {Grabowski},\ and\ \citenamefont
		{Ramdas}}]{JagannathPRB1981}%
	\BibitemOpen
	\bibfield  {author} {\bibinfo {author} {\bibfnamefont {C.}~\bibnamefont
			{Jagannath}}, \bibinfo {author} {\bibfnamefont {Z.~W.}\ \bibnamefont
			{Grabowski}},\ and\ \bibinfo {author} {\bibfnamefont {A.~K.}\ \bibnamefont
			{Ramdas}},\ }\href {https://doi.org/10.1103/PhysRevB.23.2082} {\bibfield
		{journal} {\bibinfo  {journal} {Phys. Rev. B}\ }\textbf {\bibinfo {volume}
			{23}},\ \bibinfo {pages} {2082} (\bibinfo {year} {1981})}\BibitemShut
	{NoStop}%
	\bibitem [{\citenamefont {Mayur}\ \emph {et~al.}(1993)\citenamefont {Mayur},
		\citenamefont {Sciacca}, \citenamefont {Ramdas},\ and\ \citenamefont
		{Rodriguez}}]{MayurPRB1993}%
	\BibitemOpen
	\bibfield  {author} {\bibinfo {author} {\bibfnamefont {A.~J.}\ \bibnamefont
			{Mayur}}, \bibinfo {author} {\bibfnamefont {M.~D.}\ \bibnamefont {Sciacca}},
		\bibinfo {author} {\bibfnamefont {A.~K.}\ \bibnamefont {Ramdas}},\ and\
		\bibinfo {author} {\bibfnamefont {S.}~\bibnamefont {Rodriguez}},\ }\href
	{https://doi.org/10.1103/PhysRevB.48.10893} {\bibfield  {journal} {\bibinfo
			{journal} {Phys. Rev. B}\ }\textbf {\bibinfo {volume} {48}},\ \bibinfo
		{pages} {10893} (\bibinfo {year} {1993})}\BibitemShut {NoStop}%
	\bibitem [{\citenamefont {Salfi}\ \emph {et~al.}(2014)\citenamefont {Salfi},
		\citenamefont {Mol}, \citenamefont {Rahman}, \citenamefont {Klimeck},
		\citenamefont {Simmons}, \citenamefont {Hollenberg},\ and\ \citenamefont
		{Rogge}}]{SalfiNC2014}%
	\BibitemOpen
	\bibfield  {author} {\bibinfo {author} {\bibfnamefont {J.}~\bibnamefont
			{Salfi}}, \bibinfo {author} {\bibfnamefont {J.~A.}\ \bibnamefont {Mol}},
		\bibinfo {author} {\bibfnamefont {R.}~\bibnamefont {Rahman}}, \bibinfo
		{author} {\bibfnamefont {G.}~\bibnamefont {Klimeck}}, \bibinfo {author}
		{\bibfnamefont {M.~Y.}\ \bibnamefont {Simmons}}, \bibinfo {author}
		{\bibfnamefont {L.~C.~L.}\ \bibnamefont {Hollenberg}},\ and\ \bibinfo
		{author} {\bibfnamefont {S.}~\bibnamefont {Rogge}},\ }\href
	{https://doi.org/10.1038/nmat3941} {\bibfield  {journal} {\bibinfo  {journal}
			{Nature Mater.}\ }\textbf {\bibinfo {volume} {13}},\ \bibinfo {pages} {605}
		(\bibinfo {year} {2014})}\BibitemShut {NoStop}%
	\bibitem [{\citenamefont {Koiller}\ \emph {et~al.}(2001)\citenamefont
		{Koiller}, \citenamefont {Hu},\ and\ \citenamefont
		{Das~Sarma}}]{KoillerPRL2001}%
	\BibitemOpen
	\bibfield  {author} {\bibinfo {author} {\bibfnamefont {B.}~\bibnamefont
			{Koiller}}, \bibinfo {author} {\bibfnamefont {X.}~\bibnamefont {Hu}},\ and\
		\bibinfo {author} {\bibfnamefont {S.}~\bibnamefont {Das~Sarma}},\ }\href
	{https://doi.org/10.1103/PhysRevLett.88.027903} {\bibfield  {journal}
		{\bibinfo  {journal} {Phys. Rev. Lett.}\ }\textbf {\bibinfo {volume} {88}},\
		\bibinfo {pages} {027903} (\bibinfo {year} {2001})}\BibitemShut {NoStop}%
	\bibitem [{\citenamefont {Hu}\ \emph {et~al.}(2005)\citenamefont {Hu},
		\citenamefont {Koiller},\ and\ \citenamefont {Das~Sarma}}]{XuPRB2005}%
	\BibitemOpen
	\bibfield  {author} {\bibinfo {author} {\bibfnamefont {X.}~\bibnamefont
			{Hu}}, \bibinfo {author} {\bibfnamefont {B.}~\bibnamefont {Koiller}},\ and\
		\bibinfo {author} {\bibfnamefont {S.}~\bibnamefont {Das~Sarma}},\ }\href
	{https://doi.org/10.1103/PhysRevB.71.235332} {\bibfield  {journal} {\bibinfo
			{journal} {Phys. Rev. B}\ }\textbf {\bibinfo {volume} {71}},\ \bibinfo
		{pages} {235332} (\bibinfo {year} {2005})}\BibitemShut {NoStop}%
	\bibitem [{\citenamefont {Wellard}\ and\ \citenamefont
		{Hollenberg}(2005)}]{WellardPRB2005}%
	\BibitemOpen
	\bibfield  {author} {\bibinfo {author} {\bibfnamefont {C.~J.}\ \bibnamefont
			{Wellard}}\ and\ \bibinfo {author} {\bibfnamefont {L.~C.~L.}\ \bibnamefont
			{Hollenberg}},\ }\href {https://doi.org/10.1103/PhysRevB.72.085202}
	{\bibfield  {journal} {\bibinfo  {journal} {Phys. Rev. B}\ }\textbf {\bibinfo
			{volume} {72}},\ \bibinfo {pages} {085202} (\bibinfo {year}
		{2005})}\BibitemShut {NoStop}%
	\bibitem [{\citenamefont {Li}\ \emph {et~al.}(2010)\citenamefont {Li},
		\citenamefont {Cywi\ifmmode~\acute{n}\else \'{n}\fi{}ski}, \citenamefont
		{Culcer}, \citenamefont {Hu},\ and\ \citenamefont
		{Das~Sarma}}]{QiuziPRB2010}%
	\BibitemOpen
	\bibfield  {author} {\bibinfo {author} {\bibfnamefont {Q.}~\bibnamefont
			{Li}}, \bibinfo {author} {\bibfnamefont {L.}~\bibnamefont
			{Cywi\ifmmode~\acute{n}\else \'{n}\fi{}ski}}, \bibinfo {author}
		{\bibfnamefont {D.}~\bibnamefont {Culcer}}, \bibinfo {author} {\bibfnamefont
			{X.}~\bibnamefont {Hu}},\ and\ \bibinfo {author} {\bibfnamefont
			{S.}~\bibnamefont {Das~Sarma}},\ }\href
	{https://doi.org/10.1103/PhysRevB.81.085313} {\bibfield  {journal} {\bibinfo
			{journal} {Phys. Rev. B}\ }\textbf {\bibinfo {volume} {81}},\ \bibinfo
		{pages} {085313} (\bibinfo {year} {2010})}\BibitemShut {NoStop}%
	\bibitem [{\citenamefont {Le}\ \emph {et~al.}(2017)\citenamefont {Le},
		\citenamefont {Fisher},\ and\ \citenamefont {Ginossar}}]{LePRB2017a}%
	\BibitemOpen
	\bibfield  {author} {\bibinfo {author} {\bibfnamefont {N.~H.}\ \bibnamefont
			{Le}}, \bibinfo {author} {\bibfnamefont {A.~J.}\ \bibnamefont {Fisher}},\
		and\ \bibinfo {author} {\bibfnamefont {E.}~\bibnamefont {Ginossar}},\ }\href
	{https://doi.org/10.1103/physrevb.96.245406} {\bibfield  {journal} {\bibinfo
			{journal} {Phys. Rev. B}\ }\textbf {\bibinfo {volume} {96}},\ \bibinfo
		{pages} {245406} (\bibinfo {year} {2017})}\BibitemShut {NoStop}%
	\bibitem [{\citenamefont {Dusko}\ \emph {et~al.}(2018)\citenamefont {Dusko},
		\citenamefont {Delgado}, \citenamefont {Saraiva},\ and\ \citenamefont
		{Koiller}}]{DuskoNPJQI2018}%
	\BibitemOpen
	\bibfield  {author} {\bibinfo {author} {\bibfnamefont {A.}~\bibnamefont
			{Dusko}}, \bibinfo {author} {\bibfnamefont {A.}~\bibnamefont {Delgado}},
		\bibinfo {author} {\bibfnamefont {A.}~\bibnamefont {Saraiva}},\ and\ \bibinfo
		{author} {\bibfnamefont {B.}~\bibnamefont {Koiller}},\ }\href
	{https://doi.org/10.1038/s41534-017-0051-1} {\bibfield  {journal} {\bibinfo
			{journal} {npj Quantum Inf.}\ }\textbf {\bibinfo {volume} {4}},\ \bibinfo
		{pages} {1} (\bibinfo {year} {2018})}\BibitemShut {NoStop}%
	\bibitem [{\citenamefont {Townsend}\ \emph {et~al.}(2021)\citenamefont
		{Townsend}, \citenamefont {Neuman}, \citenamefont {Debrecht}, \citenamefont
		{Aizpurua},\ and\ \citenamefont {Bryant}}]{TownsendPRB2021}%
	\BibitemOpen
	\bibfield  {author} {\bibinfo {author} {\bibfnamefont {E.}~\bibnamefont
			{Townsend}}, \bibinfo {author} {\bibfnamefont {T.}~\bibnamefont {Neuman}},
		\bibinfo {author} {\bibfnamefont {A.}~\bibnamefont {Debrecht}}, \bibinfo
		{author} {\bibfnamefont {J.}~\bibnamefont {Aizpurua}},\ and\ \bibinfo
		{author} {\bibfnamefont {G.~W.}\ \bibnamefont {Bryant}},\ }\href
	{https://doi.org/10.1103/PhysRevB.103.195429} {\bibfield  {journal} {\bibinfo
			{journal} {Phys. Rev. B}\ }\textbf {\bibinfo {volume} {103}},\ \bibinfo
		{pages} {195429} (\bibinfo {year} {2021})}\BibitemShut {NoStop}%
	\bibitem [{\citenamefont {Ashcroft}\ and\ \citenamefont
		{Mermin}(1976)}]{AshcroftBook1976}%
	\BibitemOpen
	\bibfield  {author} {\bibinfo {author} {\bibfnamefont {N.~W.}\ \bibnamefont
			{Ashcroft}}\ and\ \bibinfo {author} {\bibfnamefont {N.~D.}\ \bibnamefont
			{Mermin}},\ }\href@noop {} {\emph {\bibinfo {title} {Solid State Physics}}}\
	(\bibinfo  {publisher} {Thomson Learning},\ \bibinfo {year} {1976})\ p.\
	\bibinfo {pages} {848}\BibitemShut {NoStop}%
	\bibitem [{\citenamefont {Bardeen}(1961)}]{BardeenPRL1961}%
	\BibitemOpen
	\bibfield  {author} {\bibinfo {author} {\bibfnamefont {J.}~\bibnamefont
			{Bardeen}},\ }\href {https://doi.org/10.1103/physrevlett.6.57} {\bibfield
		{journal} {\bibinfo  {journal} {Phys. Rev. Lett.}\ }\textbf {\bibinfo
			{volume} {6}},\ \bibinfo {pages} {57} (\bibinfo {year} {1961})}\BibitemShut
	{NoStop}%
	\bibitem [{\citenamefont {Tersoff}\ and\ \citenamefont
		{Hamann}(1983)}]{TersoffPRL1983}%
	\BibitemOpen
	\bibfield  {author} {\bibinfo {author} {\bibfnamefont {J.}~\bibnamefont
			{Tersoff}}\ and\ \bibinfo {author} {\bibfnamefont {D.~R.}\ \bibnamefont
			{Hamann}},\ }\href {https://doi.org/10.1103/PhysRevLett.50.1998} {\bibfield
		{journal} {\bibinfo  {journal} {Phys. Rev. Lett.}\ }\textbf {\bibinfo
			{volume} {50}},\ \bibinfo {pages} {1998} (\bibinfo {year}
		{1983})}\BibitemShut {NoStop}%
	\bibitem [{\citenamefont {Tersoff}\ and\ \citenamefont
		{Hamann}(1985)}]{TersoffPRB1985}%
	\BibitemOpen
	\bibfield  {author} {\bibinfo {author} {\bibfnamefont {J.}~\bibnamefont
			{Tersoff}}\ and\ \bibinfo {author} {\bibfnamefont {D.~R.}\ \bibnamefont
			{Hamann}},\ }\href {https://doi.org/10.1103/PhysRevB.31.805} {\bibfield
		{journal} {\bibinfo  {journal} {Phys. Rev. B}\ }\textbf {\bibinfo {volume}
			{31}},\ \bibinfo {pages} {805} (\bibinfo {year} {1985})}\BibitemShut
	{NoStop}%
	\bibitem [{\citenamefont {Chen}(1990)}]{ChenPRB1990}%
	\BibitemOpen
	\bibfield  {author} {\bibinfo {author} {\bibfnamefont {C.~J.}\ \bibnamefont
			{Chen}},\ }\href {https://doi.org/10.1103/PhysRevB.42.8841} {\bibfield
		{journal} {\bibinfo  {journal} {Phys. Rev. B}\ }\textbf {\bibinfo {volume}
			{42}},\ \bibinfo {pages} {8841} (\bibinfo {year} {1990})}\BibitemShut
	{NoStop}%
	\bibitem [{\citenamefont {Drakova}(2001)}]{DrakovaRPP2001}%
	\BibitemOpen
	\bibfield  {author} {\bibinfo {author} {\bibfnamefont {D.}~\bibnamefont
			{Drakova}},\ }\href {https://doi.org/10.1088/0034-4885/64/2/202} {\bibfield
		{journal} {\bibinfo  {journal} {Rep. Prog. Phys.}\ }\textbf {\bibinfo
			{volume} {64}},\ \bibinfo {pages} {205} (\bibinfo {year} {2001})}\BibitemShut
	{NoStop}%
	\bibitem [{\citenamefont {Tsukada}(1994)}]{TsukadaASS1994}%
	\BibitemOpen
	\bibfield  {author} {\bibinfo {author} {\bibfnamefont {M.}~\bibnamefont
			{Tsukada}},\ }\href {https://doi.org/10.1016/0169-4332(94)90361-1} {\bibfield
		{journal} {\bibinfo  {journal} {Appl. Surf. Sci.}\ }\textbf {\bibinfo
			{volume} {76-77}},\ \bibinfo {pages} {312} (\bibinfo {year}
		{1994})}\BibitemShut {NoStop}%
	\bibitem [{\citenamefont {Ryu}\ \emph {et~al.}(2013)\citenamefont {Ryu},
		\citenamefont {Lee}, \citenamefont {Weber}, \citenamefont {Mahapatra},
		\citenamefont {Hollenberg}, \citenamefont {Simmons},\ and\ \citenamefont
		{Klimeck}}]{RyuNanoscale2013}%
	\BibitemOpen
	\bibfield  {author} {\bibinfo {author} {\bibfnamefont {H.}~\bibnamefont
			{Ryu}}, \bibinfo {author} {\bibfnamefont {S.}~\bibnamefont {Lee}}, \bibinfo
		{author} {\bibfnamefont {B.}~\bibnamefont {Weber}}, \bibinfo {author}
		{\bibfnamefont {S.}~\bibnamefont {Mahapatra}}, \bibinfo {author}
		{\bibfnamefont {L.~C.~L.}\ \bibnamefont {Hollenberg}}, \bibinfo {author}
		{\bibfnamefont {M.~Y.}\ \bibnamefont {Simmons}},\ and\ \bibinfo {author}
		{\bibfnamefont {G.}~\bibnamefont {Klimeck}},\ }\href
	{https://doi.org/10.1039/c3nr01796f} {\bibfield  {journal} {\bibinfo
			{journal} {Nanoscale}\ }\textbf {\bibinfo {volume} {5}},\ \bibinfo {pages}
		{8666} (\bibinfo {year} {2013})}\BibitemShut {NoStop}%
	\bibitem [{\citenamefont {Liebsch}(2005)}]{LiebschPRL2005}%
	\BibitemOpen
	\bibfield  {author} {\bibinfo {author} {\bibfnamefont {A.}~\bibnamefont
			{Liebsch}},\ }\href {https://doi.org/10.1103/PhysRevLett.95.116402}
	{\bibfield  {journal} {\bibinfo  {journal} {Phys. Rev. Lett.}\ }\textbf
		{\bibinfo {volume} {95}},\ \bibinfo {pages} {116402} (\bibinfo {year}
		{2005})}\BibitemShut {NoStop}%
	\bibitem [{\citenamefont {Reittu}(1995)}]{ReittuAJoP1995}%
	\BibitemOpen
	\bibfield  {author} {\bibinfo {author} {\bibfnamefont {H.~J.}\ \bibnamefont
			{Reittu}},\ }\href {https://doi.org/10.1119/1.18037} {\bibfield  {journal}
		{\bibinfo  {journal} {Am. J. Phys.}\ }\textbf {\bibinfo {volume} {63}},\
		\bibinfo {pages} {940} (\bibinfo {year} {1995})}\BibitemShut {NoStop}%
	\bibitem [{\citenamefont {Luttinger}\ and\ \citenamefont
		{Kohn}(1955)}]{LuttingerPR1995}%
	\BibitemOpen
	\bibfield  {author} {\bibinfo {author} {\bibfnamefont {J.~M.}\ \bibnamefont
			{Luttinger}}\ and\ \bibinfo {author} {\bibfnamefont {W.}~\bibnamefont
			{Kohn}},\ }\href {https://doi.org/10.1103/PhysRev.97.869} {\bibfield
		{journal} {\bibinfo  {journal} {Phys. Rev.}\ }\textbf {\bibinfo {volume}
			{97}},\ \bibinfo {pages} {869} (\bibinfo {year} {1955})}\BibitemShut
	{NoStop}%
	\bibitem [{\citenamefont {Shindo}\ and\ \citenamefont
		{Nara}(1976)}]{ShindoJPSJ1976}%
	\BibitemOpen
	\bibfield  {author} {\bibinfo {author} {\bibfnamefont {K.}~\bibnamefont
			{Shindo}}\ and\ \bibinfo {author} {\bibfnamefont {H.}~\bibnamefont {Nara}},\
	}\href {https://doi.org/10.1143/jpsj.40.1640} {\bibfield  {journal} {\bibinfo
			{journal} {J. Phys. Soc. Jpn.}\ }\textbf {\bibinfo {volume} {40}},\ \bibinfo
		{pages} {1640} (\bibinfo {year} {1976})}\BibitemShut {NoStop}%
	\bibitem [{\citenamefont {Castner}(2009)}]{CastnerPRB2009}%
	\BibitemOpen
	\bibfield  {author} {\bibinfo {author} {\bibfnamefont {T.~G.}\ \bibnamefont
			{Castner}},\ }\href {https://doi.org/10.1103/PhysRevB.79.195207} {\bibfield
		{journal} {\bibinfo  {journal} {Phys. Rev. B}\ }\textbf {\bibinfo {volume}
			{79}},\ \bibinfo {pages} {195207} (\bibinfo {year} {2009})}\BibitemShut
	{NoStop}%
	\bibitem [{\citenamefont {Greenman}\ \emph {et~al.}(2013)\citenamefont
		{Greenman}, \citenamefont {Whitley},\ and\ \citenamefont
		{Whaley}}]{GreenmanPRB2013}%
	\BibitemOpen
	\bibfield  {author} {\bibinfo {author} {\bibfnamefont {L.}~\bibnamefont
			{Greenman}}, \bibinfo {author} {\bibfnamefont {H.~D.}\ \bibnamefont
			{Whitley}},\ and\ \bibinfo {author} {\bibfnamefont {K.~B.}\ \bibnamefont
			{Whaley}},\ }\href {https://doi.org/10.1103/PhysRevB.88.165102} {\bibfield
		{journal} {\bibinfo  {journal} {Phys. Rev. B}\ }\textbf {\bibinfo {volume}
			{88}},\ \bibinfo {pages} {165102} (\bibinfo {year} {2013})}\BibitemShut
	{NoStop}%
	\bibitem [{\citenamefont {Paz}\ and\ \citenamefont
		{Soler}(2006)}]{PazPSSB2006}%
	\BibitemOpen
	\bibfield  {author} {\bibinfo {author} {\bibfnamefont {{\'{O}}.}~\bibnamefont
			{Paz}}\ and\ \bibinfo {author} {\bibfnamefont {J.~M.}\ \bibnamefont
			{Soler}},\ }\href {https://doi.org/10.1002/pssb.200541453} {\bibfield
		{journal} {\bibinfo  {journal} {Phys. Status Solidi (b)}\ }\textbf {\bibinfo
			{volume} {243}},\ \bibinfo {pages} {1080} (\bibinfo {year}
		{2006})}\BibitemShut {NoStop}%
	\bibitem [{\citenamefont {Ning}\ and\ \citenamefont {Sah}(1971)}]{NingPRB1971}%
	\BibitemOpen
	\bibfield  {author} {\bibinfo {author} {\bibfnamefont {T.~H.}\ \bibnamefont
			{Ning}}\ and\ \bibinfo {author} {\bibfnamefont {C.~T.}\ \bibnamefont {Sah}},\
	}\href {https://doi.org/10.1103/PhysRevB.4.3468} {\bibfield  {journal}
		{\bibinfo  {journal} {Phys. Rev. B}\ }\textbf {\bibinfo {volume} {4}},\
		\bibinfo {pages} {3468} (\bibinfo {year} {1971})}\BibitemShut {NoStop}%
	\bibitem [{\citenamefont {Pantelides}\ and\ \citenamefont
		{Sah}(1974)}]{PantelidesPRB1974}%
	\BibitemOpen
	\bibfield  {author} {\bibinfo {author} {\bibfnamefont {S.~T.}\ \bibnamefont
			{Pantelides}}\ and\ \bibinfo {author} {\bibfnamefont {C.~T.}\ \bibnamefont
			{Sah}},\ }\href {https://doi.org/10.1103/PhysRevB.10.621} {\bibfield
		{journal} {\bibinfo  {journal} {Phys. Rev. B}\ }\textbf {\bibinfo {volume}
			{10}},\ \bibinfo {pages} {621} (\bibinfo {year} {1974})}\BibitemShut
	{NoStop}%
	\bibitem [{\citenamefont {Lepage}(1978)}]{PeterLepageJCP1978}%
	\BibitemOpen
	\bibfield  {author} {\bibinfo {author} {\bibfnamefont {G.~P.}\ \bibnamefont
			{Lepage}},\ }\href {https://doi.org/10.1016/0021-9991(78)90004-9} {\bibfield
		{journal} {\bibinfo  {journal} {J. Comput. Phys.}\ }\textbf {\bibinfo
			{volume} {27}},\ \bibinfo {pages} {192} (\bibinfo {year} {1978})}\BibitemShut
	{NoStop}%
	\bibitem [{\citenamefont {Hahn}(2005)}]{HahnCPC2005}%
	\BibitemOpen
	\bibfield  {author} {\bibinfo {author} {\bibfnamefont {T.}~\bibnamefont
			{Hahn}},\ }\href {https://doi.org/10.1016/j.cpc.2005.01.010} {\bibfield
		{journal} {\bibinfo  {journal} {Comput. Phys. Commun.}\ }\textbf {\bibinfo
			{volume} {168}},\ \bibinfo {pages} {78} (\bibinfo {year} {2005})}\BibitemShut
	{NoStop}%
\end{thebibliography}
%

\end{document}